\documentclass[paper,notoc]{JHEP3}
\usepackage{epsfig,cite}
\usepackage{amsbsy} 
\usepackage{epsfig}
\usepackage{graphicx}

\newcommand\new{\newcommand}         

\def\beq{\begin{equation}}   
\def\eeq{\end{equation}}
\def\bea{\begin{eqnarray}}  
\def\eea{\end{eqnarray}} 

\def\eps{\epsilon}

\def\ln{\hbox{ln}}

\new{\emem}{{\ifmmode\mathrm{e}^-\else e$^-$\fi}}
\new{\epem}{{\ifmmode\mathrm{e}^+\else e$^+$\fi}}
\new{\zbo}  {{\ifmmode\mathrm{Z}\else Z\fi}}
\new{\wpm} {{\ifmmode\mathrm{W}^\pm\else W$^\pm$\fi}}
\new{\wbo} {{\ifmmode\mathrm{W}\else W\fi}}
\new{\epm} {{\ifmmode\mathrm{e^+e^-}\else $\mathrm{e^+e^-}$\fi}}
\new{\qq}  {{\ifmmode\mathrm{q}\else q\fi}}
\new{\qqb} {{\ifmmode\bar{\mathrm{q}}\else $\bar{\mathrm{q}}$\fi}}
\new{\tq}  {{\ifmmode\mathrm{t}\else t\fi}}
\new{\tqb} {{\ifmmode\bar{\mathrm{t}}\else $\bar{\mathrm{t}}$\fi}}
\new{\bq}  {{\ifmmode\mathrm{b}\else b\fi}}
\new{\bqb} {{\ifmmode\bar{\mathrm{b}}\else $\bar{\mathrm{b}}$\fi}}
\new{\ttbar}{\tq\tqb}
\new{\qqbar}{\qq\qqb}
\new{\gu}  {{\ifmmode\mathrm{g}\else g\fi}}
\new{\qqbarg}{\qq\qqb\gu}
\new{\pp}  {{\ifmmode\mathrm{p}\else p\fi}}
\new{\K}{\ensuremath{K}}
\new{\hh}  {{\ifmmode h\else $h$\fi}}
\new{\HH}  {{\ifmmode \mathrm{H}\else $\mathrm{H}$\fi}}
\new{\higgs}{\ensuremath{\mathrm{H}}}
\new{\W}{\ensuremath{\mathrm{W}}}
\new{\WW}{\W\W}
\new{\mll}{\ensuremath{m_{\ell\ell}}}
\new{\bbbar}{\ensuremath{b\bar{b}}}
\new{\ppbar}{\ensuremath{\mathrm{p}\bar{\mathrm{p}}}}
\new{\muf}{\ensuremath{\mu_\mathrm{F}}}
\new{\mur}{\ensuremath{\mu_\mathrm{R}}}
\new{\fe}  {{\ifmmode f\else $f$\fi}}
\new{\lp}  {{\ifmmode \ell\else $\ell$\fi}}
\new{\XX}  {{\ifmmode X\else $X$\fi}}
\new{\Vp}  {{\ifmmode V\else $V$\fi}}
\new{\Kzs} {{\ifmmode\mathrm{K}_\mathrm{S}^0\else $\mathrm{K}_\mathrm{S}^0$\fi}}
\new{\Kzl} {{\ifmmode\mathrm{K}_\mathrm{L}^0\else $\mathrm{K}_\mathrm{L}^0$\fi}}
\new{\Kp} {{\ifmmode\mathrm{K}\else $\mathrm{K}$\fi}}
\new{\ppHWW} {{\ifmmode\pp\pp\rightarrow\HH\rightarrow\wbo\wbo
                             \else $\pp\pp\rightarrow\HH\rightarrow\wbo\wbo$\fi}}
\new{\ppHWWlept} {{\ifmmode\pp\pp\rightarrow\HH\rightarrow\wbo\wbo\rightarrow\lp\nu\lp\nu
                             \else $\pp\pp\rightarrow\HH\rightarrow\wbo\wbo\rightarrow\lp\nu\lp\nu$\fi}}
\new{\HWWlept} {{\ifmmode\HH\rightarrow\wbo\wbo\rightarrow\lp\nu\lp\nu
                             \else $\HH\rightarrow\wbo\wbo\rightarrow\lp\nu\lp\nu$\fi}}

\new{\LEP}        {\mbox{\small\textsc{LEP}}}
\new{\LEPONE}     {\mbox{\small\textsc{LEP1}}}
\new{\LEPTWO}     {\mbox{\small\textsc{LEP2}}}
\new{\CERN}       {\mbox{\small\textsc{CERN}}}
\new{\ALEPH}      {\mbox{\small\textsc{ALEPH}}}
\new{\DELPHI}     {\mbox{\small\textsc{DELPHI}}}
\new{\LD}         {\mbox{\small\textsc{L3}}}
\new{\OPAL}       {\mbox{\small\textsc{OPAL}}}
\new{\SPS}        {\mbox{\small\textsc{SPS}}}
\new{\TEVATRON}   {\mbox{\small\textsc{TEVATRON}}}
\new{\LHC}        {\mbox{\small\textsc{LHC}}}
\new{\FERMILAB}   {\mbox{\small\textsc{FERMILAB}}}
\new{\CDF}        {\mbox{\small\textsc{CDF}}}
\new{\DZERO}      {\mbox{\small\textsc{D0}}}
\new{\CTEQ}        {\mbox{\small\textsc{CTEQ}}}
\new{\FNAL}        {\mbox{\small\textsc{FNAL}}}
\new{\ATLAS}        {\mbox{\small\textsc{ATLAS}}}
\new{\CMS}        {\mbox{\small\textsc{CMS}}}

\new{\eV}         {{\ifmmode {\mathrm{ eV}}\else ${\mathrm{ eV}}$\fi}}
\new{\MeV}        {{\ifmmode {\mathrm{ MeV}}\else ${\mathrm{ MeV}}$\fi}}
\new{\MeVc}       {{\ifmmode {\mathrm{ MeV}}/c\else ${\mathrm{ MeV}}/c$\fi}}
\new{\MeVcc}      {{\ifmmode {\mathrm{ MeV}}/c^2\else ${\mathrm{ MeV}}/c^2$\fi}}
\new{\GeV}        {{\ifmmode {\mathrm{ GeV}}\else ${\mathrm{ GeV}}$\fi}}
\new{\GeVc}       {{\ifmmode {\mathrm{ GeV}}/c\else ${\mathrm{GeV}}/c$\fi}}
\new{\GeVcc}      {{\ifmmode {\mathrm{ GeV}}/c^2\else ${\mathrm{GeV}}/c^2$\fi}}
\new{\TeV}        {{\ifmmode {\mathrm{ TeV}}\else ${\mathrm{ TeV}}$\fi}}

\new{\fb}        {{\ifmmode {\mathrm{ fb}}\else ${\mathrm{ fb}}$\fi}}
\new{\fbinv}   {{\ifmmode {\mathrm{ fb}^{-1}}\else ${\mathrm{ fb}^{-1}}$\fi}}
\new{\pb}        {{\ifmmode {\mathrm{ pb}}\else ${\mathrm{ pb}}$\fi}}
\new{\pbinv}   {{\ifmmode {\mathrm{ pb}^{-1}}\else ${\mathrm{ pb}^{-1}}$\fi}}

\new{\FEHiP}        {\mbox{\small\textsc{FEHiP}}}

\new{\Mz}         {{\ifmmode M_{\mathrm{ Z}}
                    \else $M_{\mathrm{ Z}}$\fi}}
\new{\Mzsq}       {{\ifmmode M^2_{\mathrm{ Z}}
                    \else $M^2_{\mathrm{ Z}}$\fi}}
\new{\Mw}         {{\ifmmode M_{\mathrm{ W}}
                    \else $M_{\mathrm{ W}}$\fi}}
                    
\new{\MH}         {{\ifmmode M_{\mathrm{ H}}
                    \else $M_{\mathrm{ H}}$\fi}}

\new{\as}[1]      {{\ifmmode\alpha^{#1}_s
                    \else$\alpha^{#1}_s$\fi}}
\new{\asx}[1]      {{\ifmmode a^{#1}_s
                    \else $a^{#1}_s$\fi}}
\new{\asb}[1]     {{\ifmmode\overline{\alpha}^{#1}_s
                    \else $\overline{\alpha}^{#1}_s$\fi}}
\new{\asmz}       {{\ifmmode\alpha_s(\Mzsq)
                    \else $\alpha_s(\Mzsq)$\fi}}
\new{\lqcd}       {{\ifmmode\Lambda_{\mathrm{ QCD}}
                    \else $\Lambda_{\mathrm{ QCD}}$\fi}}
\new{\lqcdsq}     {{\ifmmode\Lambda^2_{\mathrm{ QCD}}
                    \else $\Lambda^2_{\mathrm{ QCD}}$\fi}}
\new{\llla}       {{\ifmmode\Lambda_{\mathrm{ LLA}}
                    \else $\Lambda_{\mathrm{ LLA}}$\fi}} 
\new{\lmsbar}[1]  {{\ifmmode \Lambda^{(#1)}_{\overline{\mathrm{MS}}}
                    \else $\Lambda^{(#1)}_{\overline{\mathrm{MS}}}$\fi}}
\new{\lmsb}       {{\ifmmode \Lambda_{\overline{\mathrm{MS}}}
                    \else $\Lambda_{\overline{\mathrm{MS}}}$\fi}}
\new{\lmsbsq}     {{\ifmmode \Lambda^{2}_{\overline{\mathrm{MS}}}
                    \else $\Lambda^{2}_{\overline{\mathrm{MS}}}$\fi}}
\new{\pt}       {{\ifmmode p_{\mathrm{T}}
                    \else $p_{\mathrm{T}}$\fi}}
\new{\Etmiss}       {{\ifmmode E_{\mathrm{T}}^{\mathrm{miss}}
                    \else $E_{\mathrm{T}}^{\mathrm{miss}}$\fi}}
\new {\MET}{\ensuremath{E_\mathrm{T,miss}}}
\new {\METcut}{\ensuremath{E_\mathrm{T,miss}^{\mathrm{\;cut}}}}
\new {\ET}{\ensuremath{E_\mathrm{T}}}
\new {\kt}{\ensuremath{k_\mathrm{T}}}
\new{\ptlmin}       {{\ifmmode p_{\mathrm{T}}^{\lp\mathrm{min}}
                    \else $p_{\mathrm{T}}^{\lp\mathrm{min}}$\fi}}
\new{\ptlmax}       {{\ifmmode \mathrm{p}_{\mathrm{t,max}}^{\mathrm{cut}}
                    \else $\mathrm{p}_{\mathrm{t,max}}^{\mathrm{cut}}$\fi}} 
\new{\ptlep}       {{\ifmmode p_{\mathrm{T}}^{\mathrm{lepton}}
                    \else $p_{\mathrm{T}}^{\mathrm{lepton}}$\fi}}  
\new{\ptjet}       {{\ifmmode p_{\mathrm{T}}^{\mathrm{jet}}
                    \else $p_{\mathrm{T}}^{\mathrm{jet}}$\fi}}                                      
\new{\ptveto}       {{\ifmmode p_{\mathrm{T}}^{\mathrm{veto}}
                    \else $p_{\mathrm{T}}^{\mathrm{veto}}$\fi}}

\new{\HERWIG}         {\mbox{\small\textsc{HERWIG}}}
\new{\PYTHIA}         {\mbox{\small\textsc{PYTHIA}}}
\new{\JIMMY}         {\mbox{\small\textsc{JIMMY}}}

\new{\fehip}        {\mbox{\small\textsc{FEHiP}}}
\new{\fewz}        {\mbox{\small\textsc{FEWZ}}}
\new{\hqt}        {\mbox{\small\textsc{HqT}}}
\new{\mcnlo}        {\mbox{\small\textsc{MC@NLO}}}
\new{\pvegas}        {\mbox{\small\textsc{PVEGAS}}}

\newcommand{\NLmap}[2]{#1 \rightarrow \frac{#1 #2}{1- #1 + #2}}

\title{\boldmath  On the factorization of overlapping  singularities at NNLO}

\author{Charalampos Anastasiou\\
  Institute for Theoretical Physics, ETH Zurich,
  8093 Zurich, Switzerland\\
  E-mail: \email{babis@phys.ethz.ch}}

\author{Franz Herzog\\
  Institute for Theoretical Physics, ETH Zurich,
  8093 Zurich, Switzerland\\
  E-mail: \email{fherzog@itp.phys.ethz.ch}}

\author{Achilleas Lazopoulos\\
  Institute for Theoretical Physics, ETH Zurich,
  8093 Zurich, Switzerland\\
  E-mail: \email{lazopoli@itp.phys.ethz.ch}}

\abstract{ 
Real and virtual corrections in   NNLO  QCD  require multi-dimensional integrals with overlapping  singularities.   
We first review ideas and methods which have been proposed for performing   such computations.  
We then present  a new method  for the factorization of overlapping  singularities  based on non-linear integral transformations.  
We  apply this method for the evaluation of all integral topologies  which appear in double real radiation corrections in cross-section 
calculations
 for  the production of a heavy system at  hadron colliders.   
Finally, we  demonstrate with typical examples that two-loop virtual corrections  are amenable to the same method. 
} 

\begin{document}



\section{Introduction} 

Modern accelerator experiments require precise perturbative calculations for the event rate of  a variety of physical processes. 
Jets, electroweak gauge bosons and heavy quarks  are being produced  copiously at the Tevatron and the LHC. The precision of the 
measurements of physical masses, coupling parameters  and the structure of colliding  hadrons depends significantly on 
theoretical uncertainties which are better controlled at higher orders  in perturbation theory. The exclusion  of 
hypotheses for novel particles  and interactions is more significant when candidate  signal processes are predicted accurately. 
With the arrival of new discoveries,  the nature of physics laws will be deciphered more confidently with the aid of solid  quantitative 
theory predictions. 

Our abilities to simulate  complicated physical processes beyond the leading order (LO) have been improved  dramatically  in the last few years. 
At next-to-leading-order (NLO), previously inaccessible calculations with up to five particles in the  final state are  now 
possible~\cite{Berger:2010zx}. Basic collider processes with fewer particles have also been computed at 
next-to-next-to-leading (NNLO) order in QCD~\cite{Hamberg:1990np,Harlander:2002wh,Anastasiou:2002yz,Ravindran:2003um,Anastasiou:2003ds,fehip,Melnikov:2006di,GehrmannDeRidder:2008ug,GehrmannDeRidder:2007hr,Weinzierl:2008iv,Catani:2007vq,Catani:2009sm}. For hadron colliders, the 
experimental frontier in particle  physics,  only cross-sections for $2 \to 1$ processes  have been computed at NNLO. Such computations must 
be extended to $2 \to 2$ processes  which are  relevant to the  experimental program.  These include top and bottom quark pair production, 
inclusive jet production, electroweak diboson   production, electroweak gauge boson and Higgs  production in association with jets, single top 
production and beyond the Standard Model signals. It is unclear whether existing methods are suited to this  task, and   refinements of  
traditional methods or the development of new ones are  required in order to face the increased complexity of such calculations.  

A fundamental technical difficulty in NNLO calculations is the  appearance of multi-dimensional integrals 
over the momenta of up to two  additional real or virtual particles  with respect to the Born process.  These integrals  are separately infrared 
divergent  and only their sum is  finite. Higher order computations are performed almost exclusively within dimensional regularization, 
where real and  virtual corrections are expanded in  a  dimensional regulator  $\epsilon=2 -\frac{D}{2}$, where $D$ is the number of dimensions.  
Laurent expansions in $\epsilon$ are intricate in  the presence of  overlapping singularities. In this paper we present a new method for the calculation 
of the Laurent series in $\epsilon$ of multidimensional integrals which typically  appear in NNLO computations and generic  higher order computations.  

Existing methods which tackle or bypass the problem of overlapping singularities are based on 
differential equations~\cite{Gehrmann:1999as,Bern:1993kr,Kotikov:1990kg}, 
Mellin-Barnes representations~\cite{Smirnov:1999gc,Tausk:1999vh,Anastasiou:2005cb,Czakon:2005rk} 
and sector decomposition\cite{Binoth:2000ps,Hepp:1966eg,Roth:1996pd}. 
The first two approaches can be applied to the calculation of virtual or inclusive  real 
radiation corrections.  A  subtraction method can reduce the problem of fully differential real radiation 
calculations at NNLO to inclusive phase-space calculations~\cite{GehrmannDeRidder:2007hr,Catani:2007vq,Weinzierl:2008iv}
enabling  the method  of differential equations  and Mellin-Barnes for fully  differential calculations.  
Sector  decomposition can be applied universally, for virtual inclusive phase-space integrations  and fully 
differential integrations of real radiation matrix-elements. 

Sector decomposition has been the first successful method for performing fully  differential NNLO calculations for hadron 
collider processes\cite{fehip,Melnikov:2006di}.  This  is largely  attributed  to the conceptual simplicity of the method and 
its  algorithmic nature  which permits a full automatization. 
The algorithm eliminates overlapping  singularities by slicing  the integration volume such that variables which contribute 
to an overlapping  singular limit are ordered.  In this way,  the singularity is  always factorized  and it appears as a  single 
singular  limit of only the largest variable.  While this algorithm leads to numerically  stable evaluations, it  requires 
a large number of integrals  due to the slicing of the integration volume.  This  hinders  the application of the method 
to processes  with more complicated  matrix-elements.

We present here an alternative method for the factorization of overlapping singularities. We have observed that a  
factorization is possible by means of simple  rescaling of the integration variables and non-linear transformations  
which preserve the geometry of the integration boundaries. Our method   leads to a rather small number of 
numerically stable integrations. 

We apply our technique to all singular integral topologies  which appear  in the evaluation of NNLO 
double real radiation corrections to production processes of a massive system at hadron colliders. 
We  present suitable phase-space    parameterizations, analyze  the singularity structure of the matrix-elements, 
and  demonstrate how to obtain their expansion in $\epsilon$ with simple  changes of  integration variables.  
We then demonstrate how our technique  can be applied  to very complicated and maximally singular two-loop master integrals. 
In massless two-loop boxes overlapping  singularities are very hard to treat with non-linear transformations, and we  have  not been 
able to   find  suitable ones which factorize them completely. 
On the other hand,  a  hybrid  approach of  non-linear transformations combined  with sector decomposition is straightforward and more  
efficient than applying only sector  decomposition.

In Section~\ref{sec:laurent_expansion}, we  review existing methods  for the Laurent expansion in the dimensional regulator 
of integrals in higher perturbative orders. In Section~\ref{illustrative_example} we present  our method and we demonstrate  it 
on typical examples of integrals with overlapping singularities in Section~\ref{sec:examples}. In Section~\ref{sec:doublereal_general}
we discuss phase-space parameterizations and the singularity structure of  double real radiation at  NNLO. In Section~\ref{sec:RRnumerics} 
we present the numerical evaluation of integrals from all topologies  which appear in double real radiation corrections at NNLO.  
In Section~\ref{sec:loops} we  apply our method to  maximally singular  two-loop integrals, the crossed-triangle and  the crossed-box.  
Finally, we present  our conclusions in Section~\ref{sec:conclusions}.


\section{Laurent expansion  of Feynman integrals in the dimensional regulator}
\label{sec:laurent_expansion}

Loop integrals  and phase-space integrals for  the calculation of production rates of  physical processes in 
quantum field theory  are divergent in  four  space-time dimensions. In dimensional regularization, $d=4-2\epsilon$, 
all  divergent integrals are computed as a Laurent  expansion in the  dimensional regulator $\epsilon$. 
This task  is  tedious due to physical singularities, corresponding to infrared  and  collinear configurations  of 
real and virtual  particles.  Singular  manifolds in the integration volume  are of  increasing complexity at higher 
orders in  perturbation theory.  We  shall consider  examples of physical loop and phase-space integrals in later sections 
of this  paper. Here, we shall present  illustrative mathematical examples with similar  singular  behavior as in 
realistic cases. 

The easiest category of  singular integrals  is when  divergences in the integrand  occur as  poles of a single variable. 
Consider  an integral
\begin{equation}
I = \int_0^1 d^N\vec{x} \frac{ f(x_1,\ldots,x_N)}{ x_1^{1- \epsilon}}.  
\end{equation}
with $f(\vec{x})$ being a finite function in the integration volume.  This integral is  divergent for $\epsilon =0$ due  to the pole in $x_1 =0$. 
To expand  in $\epsilon$ we use a  subtraction technique, isolating the pole contribution,  
\begin{equation}
I = \int_0^1 d^N\vec{x} \frac{ f(x_1,x_2,\ldots,x_N) - f(0, x_2,\ldots,x_N) }{ x_1^{1-\epsilon}}
+ \int_0^1 d^N\vec{x} \frac{f(0, x_2,\ldots,x_N) }{ x_1^{1-\epsilon}}.
\end{equation}
In the first  term, we are allowed  to perform a Taylor 
expansion in $\epsilon$, given that the integrand is finite in the limit $x_1 \to 0$. 
In the second term, we  can perform the  integration in $\epsilon$ easily. We  then have 
\begin{eqnarray}
I &=& \int_0^1 d^N\vec{x} \frac{ f(x_1,x_2,\ldots,x_N) - f(0, x_2,\ldots,x_N) }{ x_1} \left[ 
\sum_{n=0}^\infty \frac{\epsilon^n}{n!} \log^n(x_1) \right] \nonumber \\  
&+& \frac{1}{\epsilon} \int_0^1 d^N\vec{x} f(0, x_2,\ldots,x_N).
\end{eqnarray}  
Equivalently, we  write 
\begin{eqnarray}
&& \int_0^1 d^N\vec{x}  x_1^{-1+ \epsilon} \,  f(x_1,\ldots,x_N)   = \nonumber \\ 
&&  \int_0^1 d^N\vec{x} 
\left\{ 
 \frac{\delta(x_1)}{\epsilon} +  
\sum_{n=0}^\infty \frac{\epsilon^n}{n!} \left[ \frac{\log^n(x_1)}{x_1} \right]_+
\right\}
\,  f(x_1,\ldots,x_N).  
\end{eqnarray}
All integral coefficients  in the Laurent series of  the last  expression can be evaluated  numerically.   
In case of many factorized  singularities,  
\begin{equation}
I = \int_0^1 d^N\vec{x} \frac{ f(\vec{x})}{ x_{i_1}^{1- a_1 \epsilon} \ldots x_{i_m}^{1- a_m \epsilon}},   
\end{equation}
we can apply  readily the same  procedure, and obtain a Laurent  expansion in $\epsilon$ with the substitution 
\begin{equation}
x_i^{-1+a_i \epsilon} = 
\frac{\delta(x_i)}{a_i \epsilon} +  
\sum_{n=0}^\infty \frac{a_i^n \epsilon^n}{n!} \left[ \frac{\log^n(x_i)}{x_i} \right]_+
\end{equation}
We note that one may also encounter  singularities  due to poles of second or  higher order, as for example in  
\begin{equation}
I = \int_0^1 dx \frac{f(x)}{x^{2-\epsilon}}. 
\end{equation}
The subtraction method can be also applied here, writing 
\begin{equation}
I = \int_0^1 dx \frac{f(x) - f(0) - x f^\prime(0) } {x^{2-\epsilon}} + f(0) \int_0^{1} dx x^{-2+\epsilon} + f^\prime(0) \int_0^1dx x^{-1+\epsilon}. 
\label{quadratic_singularity_expansion}
\end{equation}
The integrals in $x$ of the above expression can be computed numerically (as an expansion in $\epsilon$).  

The extraction of divergences is more complicated for integrals with overlapping singularities. 
Consider as  an example the integral  
\begin{equation}
\label{eq:basic_example}
I = \int_0^1 dx_1 dx_2  \frac{ 1}{ \left(x_1 + c x_2\right)^{2+ \epsilon}}.  
\end{equation}
For  $\epsilon =0$ the integrand becomes divergent  when both $x_1,x_2 \to 0$.  Here, it is easy to perform successively  both integrations, 
finding the  explicit result 
\begin{equation}
I = \frac{1}{\epsilon  (1+\epsilon)} \left\{ 
-c^{-1-\epsilon} + \frac{(1+c)^{-e} -1 }{c}
\right\}.
\end{equation}

\subsection{The  differential equation method}
Analogous  problems  in realistic  NNLO calculations  are  very hard to tackle with direct  analytic  integrations.  A powerful method  
which has found  numerous applications   is the method of differential equations. In this approach we find a  physical parameter for 
the integral and formulate  a differential equation using integration by parts. In our example, we can write a differential equation with respect to 
the parameter $c$, by integrating the total derivative
\begin{equation}
\int_0^1 dx_1 dx_2 \frac{\partial}{\partial x_2} \frac{x_2}{(x_1 +c x_2)^{2+\epsilon}}.
\end{equation}   
This yields the differential equation, 
\begin{equation}
\label{eq:de1}
c \frac{\partial I}{\partial c} +  I  = I_{{\rm simpler}}
\end{equation}
The inhomogeneous term on the right side of the above equation is
simpler than $I$. Specifically, 
\begin{equation}
\label{eq:de2}
I_{{\rm simpler}} = \int_0^1dx_1 \frac{1}{(c+x_1)^{2+\epsilon}}.
\end{equation}
and we find  
\begin{equation}
I_{{\rm simpler}} = \frac{-1}{1+\epsilon} \left[   
(1+c)^{-1-\epsilon} - c^{-1-\epsilon}
\right].
\end{equation}
The general solution of Eq.~\ref{eq:de1} involves integrals  over one
variable only, 
\begin{equation}
I = \frac{1}{c} \left[ {\rm const.} \, + \, \int^c d\tilde{c} \;I_{{
\rm simpler}}(\tilde{c}) \right]
\end{equation}
thus  bypassing  the problem of overlapping singularities. The
constant of integration can be determined from knowing the integral at
a special value of  $c$ or  by exploiting a  known limiting behavior
or  scaling. For  example, in our  case, we could be using that 
\begin{equation}
I(1/c) = c^{2+\epsilon} I(c), 
\end{equation}
which we  can easily derive  with a  change of variables $x_1
\leftrightarrow x_2$ in Eq.~\ref{eq:basic_example}.

\subsection{The Mellin-Barnes representation method}

Mellin-Barnes representations allow 
a straightforward Laurent expansion of Feynman integrals by using
Cauchy's theorem. Such representations  are obtained  by using the
identity, 
\begin{equation}
\label{eq:MB}
\frac{\Gamma(N)}{(A+B)^N} = \frac{1}{2 \pi  i}\int_{w_0-i\infty}^{w_0+i\infty} dw \Gamma(-w)\Gamma(N+w) A^w B^{-N-w}, 
\end{equation}  
where the contour of integration is a  straight  line  parallel to the
imaginary axis, crossing the real axis  at  a point $w_0$ such that 
the real part of the arguments of the gamma functions are positive.  
Using Eq.~\ref{eq:MB} and integrating $x_1$ and $x_2$ for the toy
example of Eq.~\ref{eq:basic_example}, we obtain the Mellin-Barnes 
representation 
\begin{equation}
\label{eq:MBex}
I = \frac{1}{2 \pi i} \int dw \frac{ \Gamma(-w) \Gamma(2+\epsilon + w)
} {\Gamma(2+\epsilon)}
\frac { \Gamma(1+w) \Gamma(-1-\epsilon-w)}{\Gamma(2+w) \Gamma(-\epsilon-w)} c^w, 
\end{equation}
where the representation is  valid (all Gamma functions  have
arguments with positive real part) if we  choose, for example,  
$\epsilon = -0.9$ and  ${\rm Re}\, w = -0.2$. Notice that we  cannot  
find any value of ${\rm Re} w$ which renders  the integral well defined 
if we  choose $\epsilon =0$. This means that the integral develops a  
pole in $\epsilon$. A Laurent  expansion can be achieved with an analytic continuation 
method, moving  the value of  $\epsilon$ from a  value that the  integral is well defined, $\epsilon =-0.9$ in our  example, 
to $\epsilon=0$ and isolating with Cauchy's  theorem the poles which arise when the arguments of $\epsilon$ dependent Gamma 
functions become zero or negative integers. In our  example, we  find that  $\Gamma(-1-\epsilon-w)=\Gamma(0)$  develops  a pole  as  
$\epsilon=-0.8$ 
(and $w=w_0$). No other pole is encountered  by  continuing the value of $\epsilon$ further to $\epsilon=0$. We  can then  write, 
\begin{equation}
\label{eq:MBac}
I = {\rm  TaylorExpand}(I)_{\epsilon =0} + {\rm Res}_{w=-1-\epsilon} (I). 
\end{equation}
 
\subsection{The subtraction method}

The differential equation method and the Mellin-Barnes  method   bypass  the problem of overlapping  singularities  
by integrating  out  Feynman parameters  and phase-space variables and generating equivalent  
representations where overlapping  singularities  cannot  occur.  
Both methods  rely on the integration volume being  well known and  free  of parameters, 
other than the space-time dimension.  This  is  the  case for  loop integrals  and inclusive phase-space  integrations.
An important  class of phase-space integrals requires parametric  boundaries which are determined according to varied
selection criteria for the  experimentally measured  observables. For  such integrals  the differential equation and Mellin-Barnes  
methods  are  not generally suitable.  

One  approach is to use a  subtraction method  in order  to map the problem of fully differential phase-space 
integrations onto a  problem of  fully inclusive  phase-space.  Consider  the toy example, 
\begin{equation}
\label{eq:PSint}
I\left[ J\right] = \int_0^1 dx_1 dx_2 \frac{J(x_1, x_2)}{(x_1+x_2)^{2+\epsilon}}, 
\end{equation}
where  the  function $J(x_1,x_2)$ plays the role of  selecting an arbitrary  subregion of  the 
integration volume according to, for example, the wishes of the experimentalists. 
Using subtraction, we can re-write 
\begin{equation} 
I\left[J\right] = \int_0^1 dx_1 dx_2 \frac{J(x_1, x_2)-J(0, 0)}{(x_1+x_2)^{2+\epsilon}}+ J(0, 0) 
\int_0^1 dx_1 dx_2 \frac{1}{(x_1+x_2)^{2+\epsilon}}.  
\end{equation}
The first  integral contains  only an integrable singularity as $\epsilon \to 0$ 
and can be computed numerically.  The second integral  is a ``fully inclusive'' integral 
and may be  computable with the Mellin-Barnes or differential equation
method.

\subsection{The sector decomposition method}

A  different approach is to use sector  decomposition.  We  divide the integration  region   according to  
the relative  magnitude of the integration variables which are  required for  the singular limit 
 (in our  example $x_1=x_2=0$), by multiplying the integrand with an appropriate 
unity, 
\begin{equation}
1 = \Theta(x_2 > x_1) + \Theta(x_1 > x_2).
\end{equation}
This gives rise to two integration domains (sectors). In the sector with $x_1 > x_2$ we  rescale  $x_2 = x_2 x_1$, and in the sector 
with $x_2 > x_1$ we  transform $x_1 = x_1 x_2$.   We  then obtain 
\begin{equation}
I\left[ J \right] = \int_0^1 dx_1 dx_2 x_1^{-1-\epsilon}\frac{J(x_1, x_1x_2)}{(1+x_2)^{2+\epsilon}}
+\int_0^1 dx_1 dx_2 x_2^{-1-\epsilon}\frac{J(x_1 x_2, x_2)}{(1+x_1)^{2+\epsilon}}.
\end{equation}
The singularities in both integrands are now factorized and a Laurent expansion  can be easily  
achieved  with a  simple subtraction. 
The method  of  sector decomposition is  suited for all types of loop and phase-space integrals. 

It is instructive to see how the method is used  on  a  physical example. Let us consider the one-loop box 
scalar integral,  
\begin{equation}
\label{eq:def_box}
I = \int 
\frac{d^d k }{i \pi^{\frac{d}{2}}} \frac{1}{k^2 (k+p_1)^2 (k+p_1+p_2)^2 (k+p_1+p_2+p_3)^2}.
\end{equation}
The corresponding Feynman parameterization reads, 
\begin{equation}
I = \int_0^1 dx_1\ldots dx_4 \delta\left( 1 - x_1-\ldots -x_4\right)
f(x_1,\ldots , x_4)
\end{equation}
with 
\begin{equation} 
f(x_1,\ldots , x_4) \equiv 
\frac{
\Gamma(2+\epsilon) 
}
{
\left[
-s x_1 x_3 -t x_2 x_4 -i0
\right]^{2+\eps}
}. 
\end{equation}
To avoid creating  poles at the upper limit of the $x_i$ integrations  we  apply first the method of 
primary-sectors~\cite{Binoth:2000ps}.
We  write 
\begin{equation}
I = \int \, dx \, \int_0^1 \left( \prod_i dx_i \right) 
\delta\left( 1 - \sum x_i \right)
f(\{x_i\})  \sum_i \delta( x_i -x) \prod _{j \ne i}  \Theta(x_i \ge x_j)
\end{equation}
We now rescale  
\begin{equation}
x_k = y_k x,
\end{equation}
 and perform the $x$ integration. This yields 
\begin{equation}
I =  \Gamma(2+\epsilon) 
\int_0^1 dy_1\ldots dy_4  \left( \sum_i y_i \right)^{2 \eps} 
\frac{ 
\sum_i \delta(1 - y_i) 
}
{
\left[
-s y_1 y_3 -t y_2 y_4 -i0
\right]^{2+\eps}
}.  
\end{equation}
All terms in the sum can be computed in exactly the same fashion.  
For convenience, although not necessary,  we  use 
the special symmetry of this problem, $y_1 \leftrightarrow  y_3$ and 
$y_2 \leftrightarrow  y_4$, and cast the integral as  
\begin{eqnarray}
\label{eq:1Lbox_primary}
I &=&
2 \, \Gamma(2+\epsilon) 
\int_0^1 dy_1 dy_2 dy_3  \left( 1 + y_1 +y_2 +y_3 \right)^{2 \eps} 
\nonumber  \\
&& \times 
\left\{
\left[
-s y_1  -t y_2 y_3 
\right]^{-2-\eps}
+ 
\left[
-t y_1  -s y_2 y_3 
\right]^{-2-\eps}
\right\}
\end{eqnarray}
We  observe that the integral becomes singular in the following instances 
\begin{equation}
y_1 =0 \mbox{ and } (y_2=0 \mbox{ or } y_3=0).
\end{equation}
We  now apply sector  decomposition  to factorize the entangled singularity structure. 
We multiply the integrand  with 
\begin{eqnarray}
1 &=& \Theta(y_2 < y_1) + \Theta(y_1 < y_2)\left( \Theta(y_1 < y_2 y_3) + \Theta(y_2 y_3 <  y_1) \right) \nonumber \\ 
   &=& \Theta(y_2 < y_1) + \Theta(y_1 < y_2 y_3) + \Theta(y_2 y_3 <  y_1 < y_2)
\end{eqnarray}
In each of the three sectors of the above  equation we  rescale  the smallest variables with respect to the large ones, 
mapping the boundaries of the  sectors to the unit cube.  Specifically, 
\begin{eqnarray}
\Theta(y_1 < y_2 y_3):&& \quad y_1 \to y_1 y_2 y_3 \\
\Theta(y_2 y_3 < y_1 < y_2):&& \quad y_1 \to y_1 y_2 \mbox{ and }  y_3 \to y_3 y_1  \\
\Theta(y_2 < y_1):&& \quad y_2 \to y_2 y_1
\end{eqnarray} 
We then obtain a representation of the one-loop box, 
\begin{equation}
I = I_1 + I_2 + I_3, 
\end{equation}
with a simple, factorized, singularity structure:
\begin{eqnarray}
\label{eq:1Lbox_sector1}
I_1 &=&
2 \, \Gamma(2+\epsilon) 
\int_0^1 dy_1 dy_2 dy_3  \left( 1 + y_1y_2 y_3 +y_2 +y_3 \right)^{2 \eps} 
\nonumber  \\
&& \times 
\left\{
\left[
-s y_1  -t 
\right]^{-2-\eps}
+ 
\left[
-t y_1  -s  
\right]^{-2-\eps}
\right\} \left( y_2 y_3\right)^{-1-\epsilon}, 
\end{eqnarray}
\begin{eqnarray}
\label{eq:1Lbox_sector2}
I_2 &=&
2 \, \Gamma(2+\epsilon) 
\int_0^1 dy_1 dy_2 dy_3  \left( 1 + y_1y_2 +y_2 +y_3y_1 \right)^{2 \eps} 
\nonumber  \\
&& \times 
\left\{
\left[
-s  -t y_3 
\right]^{-2-\eps}
+ 
\left[
-t   -s y_3  
\right]^{-2-\eps}
\right\} \left( y_2 y_1\right)^{-1-\epsilon}, 
\end{eqnarray}
\begin{eqnarray}
\label{eq:1Lbox_sector3}
I_3 &=&
2 \, \Gamma(2+\epsilon) 
\int_0^1 dy_1 dy_2 dy_3  \left( 1 + y_1 +y_2 y_1+y_3 \right)^{2 \eps} 
\nonumber  \\
&& \times 
\left\{
\left[
-s  -t y_2 y_3 
\right]^{-2-\eps}
+ 
\left[
-t   -s y_2 y_3 
\right]^{-2-\eps}
\right\} y_1^{-1-\eps}.
\end{eqnarray}
The resulting integrals $I_1,I_2,I_3$ of  sector  decomposition can all be  expanded in $\epsilon$ 
with the subtraction method.


\section{
		Factorization of singularities with non-linear  transformations
		}
		 
\label{illustrative_example}

In this section, we propose a new  method for the factorization of overlapping singularities. 
We consider again the very simple case of an overlapping  singularity
in the two-dimensional toy example integral 
\begin{equation}
\label{eq:example1}
I_1 = \int_{0}^1 dx_1 dx_2  \frac{1}{ \left( c_1 x_1 + c_2 x_2 \right)^{2+\eps}}
\label{example1}
\end{equation}
We  shall perform  a rescaling transformation
over the entire integration region, 
\begin{equation}
x_2 = \lambda x_1,  
\end{equation}
This yields, 
\begin{equation}
\label{eq:wrong_scaling}
I_1 = \int_0^1 dx_1 x_1^{-1-\eps}
\int_0^{\frac{1}{x_1}} d\lambda  \frac{1}{(c_1 + c_2 \lambda)^{2+\epsilon}}.  
\end{equation}
We notice that  there is a  factorized singularity at $x_1=0$. In this
singular point of the $x_1$ integration, the variable $\lambda$ ranges up to $+\infty$. However,
the $\lambda$  integration is convergent at  $\epsilon=0$, since the integrand scales
as $\frac{1}{\lambda^2}$ for very large values of $\lambda$.  Therefore, we could
immediately  treat the singularity at $x_1=0$ with the subtraction method
\begin{equation}
I_1 = \int_0^1 dx_1 x_1^{-1-\eps}
\int_0^\infty 
d\lambda  \frac{1}{(c_1 + c_2 \lambda)^{2+\epsilon}} 
\left[ \Theta(\lambda < \frac{1}{x_1}) - 1 \right] 
- \frac{1}{\epsilon} \int_0^\infty d\lambda  \frac{1}{(c_1 + c_2 \lambda)^{2+\epsilon}}.  
\end{equation}
We can then  evaluate numerically the integrals which are produced
after we perform an   expansion in $\epsilon$. 

Alternatively, we could perform a transformation\footnote{Integrating $\lambda$ numerically, 
with Monte Carlo methods, requires
a transformation as well, in practice, since one needs to generate $\lambda$ from some random variable 
that is produced in $\left[ 0, 1\right]$.} on $\lambda$ to bring the 
integration region back to  $\left[ 0, 1\right]$. 
\begin{equation}
\lambda = g(u), 
\end{equation}
Such a transformation maps the integration region to,  
\begin{equation}
\int_0^1 dx_1  dx_2  = \int_0^1 dx \, x \, \int_{g^{-1}(0)}^{g^{-1}(1/x)} du g^\prime(u) 
\end{equation}
It is very important to select  carefully
this transformation.
A linear mapping 
\begin{equation}
g(u) = \frac{x_1}{u}, 
\end{equation}
is clearly ineffective, since it undoes the original rescaling of $x_2
= \lambda x_1$.  
However,  non-linear mappings, such as
\begin{eqnarray}
g(u, x)  &=& \frac{u}{x+\delta (1-u)} \\ 
g(u, x)  &=& \frac{u}{x+\delta_1 (1-u)^{\delta_2}} \\ 
g(u, x)  &=& \frac{1+x}{\sqrt{ (1+2x) (1-u) +x^2}} - 1 \\
\ldots  \nonumber  
\end{eqnarray}
are effective.
For almost all practical applications in this paper we employ the mapping
\begin{equation}
g(u, x) = \frac{u}{x+\delta(1-u)}. 
\end{equation}
with $\delta$ often chosen equal to $1$.

Explicitly, the transformation  
\begin{equation}
\label{eq:transform}
x_2  =  \frac{x_1 x_2^\prime}{x_1 + (1-x_2^\prime)}, \qquad  
1-x_2  =  \frac{(1+x_1) (1-x_2^\prime)}{x_1 + (1-x_2^\prime)},
\end{equation}
with  a Jacobian 
\begin{equation}
\label{eq:jacobian}
\frac{\partial x_2}{\partial x_2^\prime} = 
\frac{x_1 (1+x_1)}{\left[ x_1 + (1-x_2^\prime)\right]^2}
\end{equation}
disentangles the overlapping  singularity, transforming the integral of eq.~\ref{example1} as 
\begin{equation}
I_1 = \int_0^1 dx_1 dx_2^\prime x_1^{-1-\eps} (1+x_1) (1-x_2^\prime +x_1)^\epsilon[c_1(1-x_2^\prime+x_1)+c_2x_2^\prime]^{-2-\epsilon} 
\end{equation}
The singularity in the limit $x_1 =0$ and $\eps =0$ can 
be  subtracted  away, and a  Laurent  series expansion around $\eps =0 $ is achieved 
 using the expansion  
\begin{equation}
\label{eq:plusexpansion}
x^{-1+ \eps} = \frac{\delta(x)}{\eps} + \sum_{n=0}^{\infty} \frac{ \epsilon^n}{n!} \left[ \frac{\ln^n(x)}{x}\right]_+.  
\end{equation}

In this approach, we have achieved to factorize the overlapping singularity with a  simple  transformation. In comparison, 
a  factorization  with  sector decomposition doubles the number of  integrals, as we have seen in the previous section. 
Economizing in the number of integrals is  even more significant for physical applications  where entanglement  of singularities 
with more variables  may take place.

Let us  now revisit the one-loop box calculation using the new method instead of sector decomposition.  
Following the ``analytical-transformation'' approach, we perform the change of  variables on the integral of Eq.~\ref{eq:1Lbox_primary} 
\begin{equation}
y_1 \to \frac{y_1 y_2 y_3}{ 1 - y_1 + y_2 y_3}.
\end{equation}
This  yields the integral 
\begin{eqnarray}
\label{eq:1Lbox_factored}
I &=&
2 \, \Gamma(2+\epsilon) 
\int_0^1 dy_1 dy_2 dy_3  (y_2 y_3)^{-1-\eps}
(1-y_1 + y_2 y_3)^{-\eps}
\left[y_1 y_2 y_3 + (1+y_2+y_3) (1-y_1 + y_2 y_3) \right]^{2\eps}
\nonumber  \\
&& \times 
\left\{
\left[
-s y_1  -t (1-y_1) - t y_2 y_3 
\right]^{-2-\eps}
+ 
\left[
-t y_1  -s (1-y_1) - s y_2 y_3 
\right]^{-2-\eps}
\right\}
\end{eqnarray}
In this integral, the singularities have nicely factorized  in the term  $(y_2 y_3)^{-1-\eps}$. 
In comparison to sector decomposition, we now have to perform one integration rather than three.


\section{ 
		Characteristic  forms of entangled singularities 
		}

\label{sec:examples}

In this section, we present some typical examples of integrals with
overlapping  singularities and the mappings that we use to disentangle
them.

Our second example is the integral
\beq
I_2=\int_0^1{dx\,dy\,dz\, 
			\frac{1}
				{(x+yz)^{2+\epsilon} }
		 }
\label{example2}
\eeq
which is a trivial extension of eq.~\ref{example1}. We use the mapping 
\beq
(x+yz)\;\;\;\;\;\; :\;\;\;\;\;\; \NLmap{x}{yz} 
\label{mapping2}
\eeq
where we have also designated the singularity structure of the integral. This mapping  leads to 
\beq
I_2=\int_0^1 \frac{dx\,dy\,dz\,}
				{(yz)^{1+\epsilon}}
				\frac{(1-x+yz)^{\epsilon}}{(1+yz)^{1+\epsilon}}		 
\eeq
where the singularities are factorized in terms of $y$ and $z$.


Next, let's consider 
\beq
I_3=\int_0^1dx\,dy\,dz \frac{1}{(x+y+z)^{3+\epsilon}}.
\label{example3}
\eeq
Here we use the simultaneous double mapping
\beq
(x+y+z) \;\;\;\;\;\; :\;\;\;\;\;\; \NLmap{y}{x} \;\;,\;\; \NLmap{z}{x} 
\label{mapping3}
\eeq
which leads to 
\beq
I_3=\int_0^1\frac{dx\,dy\,dz}{x^{1+\epsilon}}\frac{(1+x)^2(1-y+x)^{1+\epsilon}(1-z+x)^{1+\epsilon}}{((1+x)^2-zy)^{3+\epsilon}}.
\eeq


Next, let's consider the integral
\beq
I_4=\int_0^1 dx\,dy\,dz\,dw \frac{1}{(x+y(z+w))^{2+\epsilon}}.
\label{example4}
\eeq
Here we use successively
\beq
(x+y(z+w)) \;\;\;\;\;\; :\;\;\;\;\;\; \NLmap{z}{w} \;\; ,\;\; \NLmap{x}{yw}.
\label{mapping4}
\eeq
The integral then becomes
\beq
I_4=\int_0^1\frac{dx\,dy\,dz\,dw}{y^{1+\epsilon}}
			\frac{(1+w)(1+yw)(1-x+yw)^\epsilon (1-z+w)^\epsilon}
				{ \left[ 1-zx+w(1+y+yw)\right]^{2+\epsilon}}.
\eeq
It is maybe instructive to see how this integral of eq.~\ref{example4} behaves under simple rescaling. Consider $z=\lambda_z w$ and then $x=\lambda_x yw$. We get 
\beq
I_4=\int_0^1\frac{dy\;dw}{y^{1+\epsilon}}\int_0^{1/w}d\lambda_z\int_0^{1/yw}d\lambda_x 
\frac{1}{(\lambda_x+\lambda_z+1)^{2+\epsilon}}.
\eeq
Only the integral over $\lambda_x$ extends to infinity and in that limit the behavior of the integrand is $d\lambda_x/\lambda_x^2$ which vanishes at infinity. 

Next let's consider 
\beq
I_5=\int_0^1 dx\,dy\,dz\,dw \frac{(xyzw)^\epsilon}{(x+y+zw)^3}.
\label{example5}
\eeq
Here we use the successive mappings 
\beq
(x+y+zw) \;\;\;\;\;\; :\;\;\;\;\;\; \NLmap{x}{zw} \;\; , \;\; \NLmap{y}{zw} 
\label{mapping5}
\eeq
which brings the integral to the factorized form 
\beq
I_5=\int_0^1 dx\,dy\,dz\,dw \frac{(xy)^\epsilon}{z^{1-3\epsilon}w^{1-3\epsilon}}
					\frac{(1+zw)^2(1-x+zw)^{1-\epsilon}(1-y+zw)^{1-\epsilon}}{(1-xy+2zw+z^2w^2)^3}.
\eeq

When overlapping singularities appear together with factorized singularities in the same variable, a slight complication appears. 		
Consider the integral
\beq
I_6=\int_0^1 dx dy \frac{(xy)^\eps}{x(x+y)}.
\eeq
It has a factorized singularity at $x=0$ and an overlapping singularity at $x=0=y$. 
Let us call the singularity at $x=0$ active and the one at $y=0$ passive.
In order to disentangle the singularity we would like to use the same mapping as in the previous examples. 
It turns out that we can do this, but only as long as we choose to remap the active singularity, i.e.
\beq
x(x+y)  \;\;\;\;\;\; :\;\;\;\;\;\;  \NLmap{x}{y}. 
\eeq
The integral then becomes 
\beq
I_6=\int_0^1 dx dy x^{-1+\epsilon} y^{-1+2\epsilon} (1-x+y)^{-\epsilon}.
\eeq
which can be subtracted easily.

Note that applying the wrong rescaling $y=\lambda x $ one gets 
\beq
I_6=\int_0^1 dx \int_0^{1/x}d\lambda \frac{(\lambda)^\epsilon}{x^{1-2\epsilon}(1+\lambda)}.
\eeq
We can see immediately that the $d\lambda$ integral is logarithmically divergent at the active singularity $x\to 0$ due to the upper limit of the integration region. On the contrary, applying the correct rescaling $x=\lambda y$ one gets
\beq
I_6=\int_0^1 dy \int_0^{1/y}d\lambda \frac{(\lambda)^\epsilon}{y^{1-2\epsilon}\lambda(1+\lambda)}
\eeq
which, as  $y\to 0$, behaves as $d\lambda /\lambda^2$ which is finite.  

We see that the simple $\lambda$-rescaling works as a guideline, showing when a mapping properly factorizes the singularities of an integral.  It will be instrumental in more complicated cases presented below.

Let us now consider the integral\footnote{We find similar singularity structures in double real radiation.}
\beq
I_7=\int_0^1 dx dy dz \frac{(xyz)^\eps}{xy(xy+z)}.
\eeq
Identifying $x=0=z$ and $y=0=z$ as two independent overlapping singularities, where $z$ is passive and both
$x$ and $y$ are active, we know from the previous example that we should not map $z$ from the previous example. 
So one is left to map $x\to z$ or $y \to z$. 
By the symmetry of the integrand it does not matter which one of these one can choose.
Let us choose $x$ and a slightly modified mapping that keeps the expressions simpler:
\beq
xy(y+z) \;\;\;\;\;\; :\;\;\;\;\;\;   x\rightarrow \frac{xz}{1-x+zx}. 
\eeq
The integrand then becomes
\beq
I_7=\int_0^1 dx dy dz \frac{(xy^2z)^\eps(1-x+zx)^{-\eps}}{xyz(x(y+z)+(1-x))}. 
\eeq
We have now managed to ``activate'' the singularity at $z=0$. At the same time the singularities
at $y=0$ and at $x=0$ have remained active as well. 
However, we now find an overlapping singularity at $z=0=y$ \emph{and} $x=1$, where the singularity at $x=1$ is passive. 
Notice that we started with two independent (partially interfering) overlapping singularities,  have treated one of them and are now left 
with only one, which lies at a different  point.
We shall remap $z$ and $y$ as follows
\beq
y\rightarrow \frac{y(1-x)}{1-y+(1-x)y}
\;\;\;\; , \;\;\;\;
z\rightarrow \frac{z(1-x)}{1-z+(1-x)z}.
\eeq
The integrand becomes
\beq
I_7=\int_0^1 dx dy dz  \frac{(xy)^{-1+\epsilon} ((1-x)z)^{-1+2\epsilon}}{(1-xy)^\epsilon (1-xz)^\epsilon (1-x^2yz)}. 
\eeq
Note that the remaining singularity of the integrand is integrable. 

Let's now explore the potential of a slightly different kind of mapping. We have the integral
\beq
I=\int_0^1\prod_i dy_i \frac	{dx}
			{ (ax+b)^{N}}
\eeq
with $a,b$ independent of $x$ but potentially dependent on $y_i$. In the latter case the integral might have
overlapping singularities, as $a,b\rightarrow 0$ or $x,b\rightarrow 0$. We employ
\beq
\NLmap{x}{b/a}
\label{map2}
\eeq
and get 
\beq
I=\int_0^1 \prod_i dy_i  \frac{ dx  (a(1-x)+b)^{N-2}} {b^{N-1} (a+b)^{N-1}}.
\eeq
If $N\geq 2 $ this mapping factorizes the singularity at $b\rightarrow 0$ and, at the same time, exposes the $a+b$ structure of the overlapping 
$a,b\rightarrow 0$ singularity, making it ready for further mappings.

Let's see, as an example, 
\beq
I_8 = \int_0^1 \frac	{dx_1\,dx_2\,dx_3\,dx_4\,dx_5}
				{
				(x_1+x_2x_3+x_2x_4+x_4x_5)^{3+\epsilon}
				}. 
\label{example_xsd_1}
\eeq
We map:
\beq
\NLmap{x_1}{(x_2x_3+x_2x_4+x_4x_5)} 
\eeq
to get 
\beq
I_8 = \int_0^1 	dx_1\,dx_2\,dx_3\,dx_4\,dx_5
		\frac{(1-x_1+x_2x_3+x_2x_4+x_4x_5)^{1+\epsilon}}
				{
				(x_2(x_3+x_4)+x_4x_5)^{2+\epsilon} (1+x_2x_3+x_2x_4+x_4x_5)^{2+\epsilon}
				} .
\eeq
We can now use the mapping of eq.~\ref{map2} with $a=x_3+x_4$ and $b=x_4x_5$ to get 
\beq
I_8=\int_0^1dx_1\,dx_2\,dx_3\,dx_4\,dx_5 \frac{F(x_i)}{x_4^{1+\epsilon}x_5^{1+\epsilon} (x_3+x_4+x_4x_5)^{1+\epsilon}}
\eeq
where $F(x_i)$ is a finite function of $x_i$. Noting that $x_4$ is an active singularity, we use the mapping of eq.~\ref{map2} again with 
$a=1+x_5$ and $b=x_3$ to get
\beq
I_8=\int_0^1 dx_1\,dx_2\,dx_3\,dx_4\,dx_5 \frac{F^\prime(x_i)}{x_4^{1+\epsilon}x_5^{1+\epsilon}x_3^{1+\epsilon}}.
\eeq

Let us now see some examples where  we employ a hybrid  method  of  one-step of sector decomposition and non-linear transformations 
to factorize  overlapping singularities. A  similar singularity structure  appears  in two-loop massless  box integrals. 


We consider
\beq
I_9=\int_0^1{
			\frac{dx_1\,dx_2\,dx_3\,dx_4\,dx_5 }
				{\left[x_1x_3+x_1x_2+x_2(x_4+x_5+x_3x_4x_5)\right]^{3+\epsilon}}
		}.
\label{example_sd_2}
\eeq
We split this integral in two sectors
\beq
x_2>x_3\;\;\;\;:\;\;\;\; \int_0^1{
				 \frac{1}{x_2^{2+\epsilon}}\frac{dx_1\,dx_2\,dx_3\,dx_4\,dx_5 }
				{\left[x_1(1+x_3)+x_4+x_5+x_3x_4x_5\right]^{3+\epsilon}}
			}
\eeq
which has the singularity structure of eq.~\ref{example3} and can be factorized by the mapping of eq.~\ref{mapping3},
and 
\beq
x_3>x_2\;\;\;\;:\;\;\;\; \int_0^1{
				 \frac{1}{x_3^{2+\epsilon}} \frac{dx_1\,dx_2\,dx_3\,dx_4\,dx_5}
				{\left[x_1(1+x_2)+x_2(x_4+x_5+x_4x_5)\right]^{3+\epsilon}}
			}
\eeq
which is of the type of eq.~\ref{example4} and can be factorized with the mapping of eq.~\ref{mapping4}.


We now move to the most complicated example of this section, the integral
\beq
I_{10}=\int_0^1{  
			\frac{dx_1dx_3dx_4 d\tau_1d\tau_2}
				{[x_1B+x_1x_3A+x_4\tau_1C+x_4\tau_2 D+x_3\tau_1\tau_2E ]^{3+2\epsilon}}
		 }
\label{example_sd_3}
\eeq
with $A,E$ finite and $B,C,D$ finite functions of $\tau_{1,2}$.

We split this integral in two sectors, $x_1,x_4$, and we get
\beq
x_1>x_4 \;\;\;\;:\;\;\;\; I_{10A}= \int_0^1{  	
			\frac{dx_1dx_3dx_4 d\tau_1d\tau_2 \;x_1}
				{[x_1(B+Ax_3+Cx_4\tau_1+Dx_4\tau_2) +Ex_3\tau_1\tau_2]^{3+2\epsilon}}
		 }
\eeq
which is of the type of eq.~\ref{example2} and can be immediately factorized with the mapping eq.~\ref{mapping2}. The other sector is 
\beq
x_4>x_1 \;\;\;\;:\;\;\;\;  I_{10B}= \int_0^1{  	
			\frac{dx_1dx_3dx_4 d\tau_1d\tau_2 \;x_4}
				{[Bx_1x_4+Ax_1x_4x_3+Cx_4\tau_1+Dx_4\tau_2 +Ex_3\tau_1\tau_2]^{3+2\epsilon}}
		 }.
\eeq
This should be further split in $x_3,x_4$ to get 
\beq
x_4>x_3 \;\;\;\;:\;\;\;\; I_{10B1}= \int_0^1{  
		 \frac{dx_1dx_3dx_4 d\tau_1d\tau_2}{x_4^{1+2\epsilon}}
			\frac{1}
				{[Bx_1+Ax_1x_4x_3+C\tau_1+D\tau_2 +Ex_3\tau_1\tau_2]^{3+2\epsilon}}
		 }
\eeq
which is of the type of eq.~\ref{example3} and we can use the mapping eq.~\ref{mapping3} to factorize it.
The other sector is 
\beq
x_3>x_4 \;\;\;\;:\;\;\;\;  I_{10B2}=\int_0^1{  
		 \frac{dx_1dx_3dx_4 d\tau_1d\tau_2}{x_3^{1+2\epsilon}}
			\frac{x_4}
				{[Bx_1x_4+Ax_1x_4x_3+Cx_4\tau_1+Dx_4\tau_2 +E\tau_1\tau_2]^{3+2\epsilon}}
		 }
\eeq
and requires further splitting. We choose to split in the variables $x_4,\tau_1$ to get 
\beq
x_4>\tau_1 \;\;\;\;:\;\;\;\;  I_{10B2A}=\int_0^1{  
		\frac{dx_1dx_3dx_4 d\tau_1d\tau_2}{x_3^{1+2\epsilon}x_4^{1+2\epsilon}}
			\frac{1}
				{[B^\prime x_1+Ax_1x_3+Cx_4\tau_1+D^\prime \tau_2 +E\tau_1\tau_2]^{3+2\epsilon}}
		 }
\eeq
which is of the type of eq.~\ref{example5}, and 
\beq
\tau_1>x_4 \;\;\;\;:\;\;\;\;  I_{10B2B}=\int_0^1{  
		\frac{dx_1dx_3dx_4 d\tau_1d\tau_2 }{x_3^{1+2\epsilon}\tau_1^{1+2\epsilon}}
			\frac{x_4}
				{[Bx_1x_4+Ax_1x_4x_3+Cx_4\tau_1+Dx_4\tau_2 +E\tau_2]^{3+2\epsilon}}
		 }
\eeq
which requires a final split in $x_4,\tau_2$,
\beq
\tau_2>x_4 \;\;\;\;:\;\;\;\;  I_{10B2B1}=\int_0^1{  
		\frac{dx_1 dx_3 dx_4  d\tau_1 d\tau_2 }
			{x_3^{1+2\epsilon}\tau_1^{1+2\epsilon} \tau_2^{1+2\epsilon}}
			\frac{x_4}
				{[Bx_1x_4+Ax_1x_4x_3+Cx_4\tau_1+Dx_4\tau_2 +E]^{3+2\epsilon}}
		 }
\eeq
which is finite and 
\beq
x_4>\tau_2 \;\;\;\;:\;\;\;\;  I_{10B2B2}=\int_0^1{  
		 \frac{dx_1dx_3dx_4 d\tau_1d\tau_2}{x_3^{1+2\epsilon}\tau_1^{1+2\epsilon}x_4^{1+2\epsilon}}
			\frac{1}
				{[Bx_1+Ax_1x_3+C\tau_1+Dx_4\tau_2 +E\tau_2]^{3+2\epsilon}}
		 }
\eeq
which is of the type of eq.~\ref{example4}. The original integral can be written in terms of its five sectors as 
\beq
I_{10}=I_{10A}+I_{10B1}+I_{10B2A}+I_{10B2B1}+I_{10B2B2}.
\eeq

Finally, let's consider 
 \beq
I_{11}=\int_0^1{  
			\frac{dx_1dx_2dx_4 d\tau_1d\tau_2\;\;x_2^{1+\epsilon}}
				{[x_1A+x_1x_2B_1B_2+x_2x_4\tau_1B_2C+x_2x_4\tau_2 B_1D+x_2\tau_1\tau_2E ]^{3+2\epsilon}}
		 }
\label{example_sd_4}
\eeq
with $A,E,C,D$ finite and $B_{1,2}=2-\tau_{1,2}$, also finite for all values of $\tau_{1,2}$. 
We begin by splitting the integral in $\tau_1$,$\tau_2$. We get 
\beq
\tau_1>\tau_2\;\;\;\; :\;\;\;\; I_{11A}=\int_0^1{  
			\frac{dx_1dx_2dx_4 d\tau_1d\tau_2\;\;x_2^{1+\epsilon}}
				{[x_1A+x_1x_2B_1B_{12}+x_2x_4\tau_1B_{12}C+x_2x_4\tau_2\tau_1 B_1D+x_2\tau_1^2\tau_2E ]^{3+2\epsilon}}
		 }
\eeq 
and
\beq
\tau_2>\tau_1\;\;\;\; :\;\;\;\; I_{11B}=\int_0^1{  
			\frac{dx_1dx_2dx_4 d\tau_1d\tau_2\;\;x_2^{1+\epsilon}}
				{[x_1A+x_1x_2B_2B_{12}+x_2x_4\tau_1\tau_2 B_{1}C+x_2x_4\tau_1 B_{12}D+x_2\tau_1\tau_2^2E ]^{3+2\epsilon}}
		 }
\eeq 
where $B_{12}=2-\tau_1\tau_2$. We notice that we can get $I_{11B}$ from $I_{11A}$ if we exchange $C$ and $D$ and rename the dummy integration variables  $\tau_1 \leftrightarrow \tau_2$, so that 
\beq
I_{11B}(C,D)= I_{11A}(D,C).
\eeq
We will use the same decompositions and mappings to factorize $I_{11B}$ and $I_{11A}$ (with $\tau_1$ and $\tau_2$ interchanged), so we only describe the latter below.
We split $I_{11A}$ in two sectors with respect to $\tau_1$ and $x_4$. We get 
\beq
\tau_1>x_4\;\;\;\; :\;\;\;\; I_{11A1}=\int_0^1{  
			\frac{dx_1dx_2dx_4 d\tau_1d\tau_2\;\;x_2^{1+\epsilon} \tau_1}
				{[x_1(A+x_2B_1B_{12})+x_2\tau_1^2(
					x_4B_{12}C+x_4\tau_2 B_1D+\tau_2E) ]^{3+2\epsilon}}
		 }.
\eeq 
We perform the mappings
\beq
 \NLmap{\tau_2}{x_4}
\eeq
and then 
\beq
\NLmap{x_1}{\tau_1^2\tau_2x_4}
\eeq
which factorize all singularities.
We also have 
\beq
x_4>\tau_1\;\;\;\; :\;\;\;\; I_{11A2}=\int_0^1{  
			\frac{dx_1dx_2dx_4 d\tau_1d\tau_2\;\;x_2^{1+\epsilon} x_4}
				{[	x_1(A+x_2B_1^\prime B_{12}^\prime)
					+x_2x_4^2\tau_1
						(B_{12}^\prime C
					+\tau_2 B_1^\prime D
					+\tau_1\tau_2E)
					 ]^{3+2\epsilon}}
		 }.
\eeq  
We will split with respect to $\tau_1$,$\tau_2$ to get 
\beq
\tau_1>\tau_2\;\;\;\; :\;\;\;\; I_{11A2a}=\int_0^1{  
			\frac{dx_1dx_2dx_4 d\tau_1d\tau_2\;\;x_2^{1+\epsilon} x_4\tau_1}
				{[	x_1(A+x_2B_1^\prime B_{12}^{\prime\prime})
					+x_2x_4^2\tau_1
						(B_{12}^{\prime\prime} C
					+\tau_2\tau_1 B_1^\prime D
					+\tau_1^2\tau_2E)
					 ]^{3+2\epsilon}}
					 }
\eeq 
which can be factorized by 
\beq
\NLmap{x_1}{x_2x_4^2\tau_1}
\eeq
and 
\beq
\tau_2>\tau_1\;\;\;\; :\;\;\;\; I_{11A2b}=\int_0^1{  
			\frac{dx_1dx_2dx_4 d\tau_1d\tau_2\;\;x_2^{1+\epsilon} x_4\tau_2}
				{[x_1(A+x_2B_1^{\prime \prime\prime} B_{12}^{\prime \prime\prime})+x_2x_4^2\tau_1\tau_2
					(B_{12}^{\prime \prime\prime} C+\tau_2  B_1^{\prime \prime\prime} D+\tau_1\tau_2^2E) ]^{3+2\epsilon}}
		 }
\eeq  
which can be factorized by 
\beq
\NLmap{x_1}{x_2x_4^2\tau_1\tau_2}.
\eeq
The original integral can, therefore, be factorized in six different integrals:  
\beq
I_{11}=I_{11A}+I_{11B}=I_{11A1}+I_{11A2a}+I_{11A2b}+(u\leftrightarrow t)
\eeq

\section{Double real radiation for  final states with massive particles}
\label{sec:doublereal_general}

One of the major challenges at NNLO in QCD has been the computation of the double real emission part of the cross-section.
While the computation of the matrix elements with $N+2$ particles in the final state is not a problem per se, difficulties arise 
when one integrates over the phase space of the two unresolved particles. The corresponding integrals 
are infrared divergent in the soft and collinear limits and are  dimensionally regulated. The divergences have to be subtracted 
before the integrals can be numerically evaluated. As long as the singularities are factorized,  as they usually are at NLO, 
it is straightforward to use a  Laurent expansion over the singular variables, and evaluate its coefficients numerically.
At NNLO, the singularity structure of the integral is more intricate, as  line and overlapping singularities appear, and 
the desired factorization is not straightforward.

The method of sector decomposition has already been applied successfully to achieve this factorization 
for hadron collider~\cite{fehip,Melnikov:2006di} and decay processes~\cite{Melnikov:2008qs,Anastasiou:2005pn,Asatrian:2006ph}.
A drawback of the method is that it leads to a large number of sectors.
The goal of this paper is to replace sector decomposition for  double-real radiation integrals  with  an economical factorization 
method based on non-linear transformations.

\subsection{Infrared singularities in double real radiation}

\begin{figure}[htb!]
\centering%
\epsfig{file=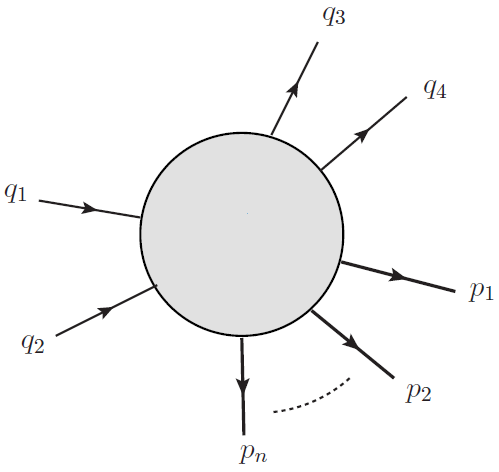,width=5cm}
\caption{Notational setup: $q_i$ are massless particles, while $p_i$ are massive.}
\label{fig:DRsetup}
\end{figure}

We consider double real emission to a generic NNLO $2\to n+2$ process 
(see Fig  \ref{fig:DRsetup}) with $n$ massive particles and 2  massless  partons in the final state. 
We denote the momenta of the incoming particles by $q_1,q_2$, those of the outgoing massive  
particles by  $p_1..p_n$ and those of the two unresolved partons by $q_3$ and $q_4$.
Infrared singularities in this phase space will occur whenever $q_3$ and/or $q_4$ become soft or collinear to $q_1$, $q_2$ 
or to each other.
For the case of double real radiation to the production of a single massive particle (e.g. Higgs, W or Z production)
potentially singular propagators can be summarized as
\begin{eqnarray}
s_{34} &=& 2q_3.q_4   \nonumber \\
s_{13} &=& -2q_1.q_3  \nonumber \\  
s_{23} &=& -2q_2.q_3  \nonumber \\
s_{14} &=& -2q_1.q_4  \nonumber \\  
s_{24} &=& -2q_2.q_4  
\end{eqnarray}
and
\begin{eqnarray}
s_{134} & = & (q_3+q_4-q_1)^2=s_{34}+s_{13}+s_{14}\nonumber \\ 
s_{234} & = & (q_3+q_4-q_2)^2=s_{34}+s_{23}+s_{24}.
\end{eqnarray}
Note that $s_{123}$, $s_{124}$ are bounded from below. 
Further soft singularities can be found if there are colored massive particles in the final state,
which can radiate off soft gluons. One can then get also the following possibly singular denominators:
\begin{eqnarray}
t_{3i} &=& 2q_3.p_i  \nonumber \\  
t_{4i} &=& 2q_4.p_i   
\end{eqnarray}
and
\begin{equation}
t_{34i}=(q_3+q_4+p_k)^2-m_k^2=s_{34}+t_{3i}+t_{4i} 
\end{equation}
for $i\geq 1$.
Since 
\begin{equation}
t_{3i}=2q_3.p_i= 2E_3(E_i-|\textbf{p}_i|\cos\theta_{3i}), \qquad E_i>|\textbf {p}_i|
\end{equation} 
the soft singularity is factorized in $E_3$. 
Whenever some heavy colored state radiates off two gluons we can also get
the denominator $t_{34i}$, it can only become singular in the double soft limit when $E_3=0=E_4$. 
However the double soft limit will always be factorized as we will show in the next section.

\begin{figure}[htb!]
\centering%
\epsfig{file=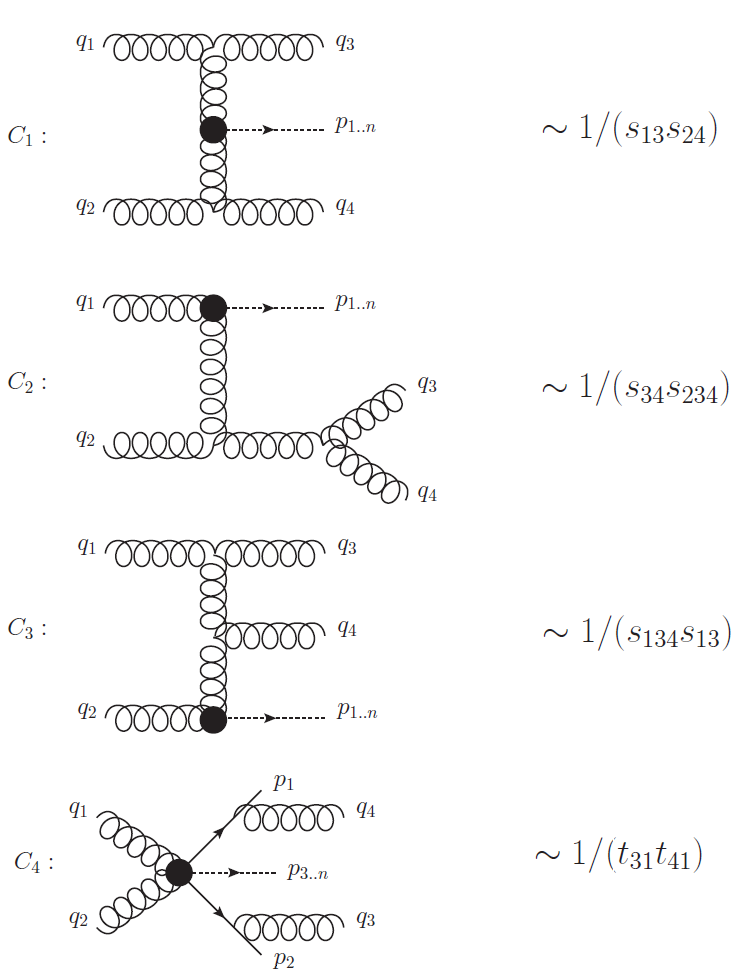,width=10cm}
\caption{Most singular topologies:$C_1,C_2,C_3,C_4$}
\label{fig:singtops}
\end{figure}

Let us now discuss the denominator structure of the most singular diagrams which 
one could expect in double real radiation: those where radiation is emitted by initial state particles. We have illustrated the propagator structure of these topologies using some diagrams containing gluons in Fig.\ref{fig:singtops}
(diagrams containing just massless quarks correspond to the same topologies).

Diagrams whose propagator structure can be related to the ones in Fig.\ref{fig:singtops} 
by a simple interchange of $q_3$ with $q_4$ or of $q_1$ with $q_2$ will also fall into the same 
topology. 

By considering square and interference terms of the topologies $C_1,C_2$ and $C_3$, 
we obtain the following list of integrals:
\begin{enumerate}
 \item Topology  $C_1 \otimes C_1$:
  \begin{equation}
  \int \frac{d\Phi_3 N(\left\lbrace s_{ij}\right\rbrace)}{(s_{13}s_{24})^2},   
  \int \frac{d\Phi_3 N(\left\lbrace s_{ij}\right\rbrace)}{s_{13}s_{23}s_{14}s_{24}} 
  \end{equation}

 \item Topology $C_2 \otimes C_2$:
  \begin{equation}
  \int \frac{d\Phi_3 N(\left\lbrace s_{ij}\right\rbrace)}{(s_{34}s_{134})^2},   
  \int \frac{d\Phi_3 N(\left\lbrace s_{ij}\right\rbrace)}{s_{34}^2 s_{134}s_{234}} 
  \end{equation}
 \item Topology $C_3 \otimes C_3$:
  \begin{equation}
  \int \frac{d\Phi_3 N(\left\lbrace s_{ij}\right\rbrace)}{(s_{13}s_{134})^2},   
  \int \frac{d\Phi_3 N(\left\lbrace s_{ij}\right\rbrace)}{s_{13}s_{23}s_{134}s_{234}}, 
  \int \frac{d\Phi_3 N(\left\lbrace s_{ij}\right\rbrace)}{s_{13}s_{24}s_{134}s_{234}} 
  \end{equation}
 \item Topology $C_1 \otimes C_2$:
  \begin{equation}
  \int \frac{d\Phi_3 N(\left\lbrace s_{ij}\right\rbrace)}{s_{34}s_{234}s_{13}s_{24}}   
  \end{equation}
 \item Topology $C_1 \otimes C_3$:
  \begin{equation}
  \int \frac{d\Phi_3 N(\left\lbrace s_{ij}\right\rbrace)}{s_{134}s_{13}s_{23}s_{14}},   
  \int \frac{d\Phi_3 N(\left\lbrace s_{ij}\right\rbrace)}{s_{134}s_{13}^2 s_{14}}  
  \end{equation}
 \item Topology $C_2 \otimes C_3$:
  \begin{equation}
  \int \frac{d\Phi_3 N(\left\lbrace s_{ij}\right\rbrace)}{s_{34}s_{134}^2s_{13}},   
  \int \frac{d\Phi_3 N(\left\lbrace s_{ij}\right\rbrace)}{s_{34}s_{134}s_{234}s_{23}}  
  \end{equation}
  \item Topology $C_4\otimes C_4$: 
  \begin{equation}
  \int \frac{d\Phi_3 N(\left\lbrace s_{ij}\right\rbrace)}{t_{i3}^2 t_{j4}^2},   
  \int \frac{d\Phi_3 N(\left\lbrace s_{ij}\right\rbrace)}{t_{i3} t_{j4}t_{j3} t_{i4}}  
  \end{equation}
  \item Topology $C_4\otimes C_1$: 
  \begin{equation}
  \int \frac{d\Phi_3 N(\left\lbrace s_{ij}\right\rbrace)}{t_{i3} t_{j4} s_{13}s_{14}}
  \end{equation}
  \item Topology $C_4\otimes C_2$: 
  \begin{equation}
  \int \frac{d\Phi_3 N(\left\lbrace s_{ij}\right\rbrace)}{t_{i3} t_{j4} s_{34}s_{134}}
  \end{equation}
  \item Topology $C_4\otimes C_3$: 
  \begin{equation}
  \int \frac{d\Phi_3 N(\left\lbrace s_{ij}\right\rbrace)}{t_{i3} t_{j4} s_{13}s_{134}}
    \end{equation}
\end{enumerate}
Where $d\Phi_3$ is the differential double emission phase space element for $2+n$ final state particles,  and 
$N(\left\lbrace s_{ij}\right\rbrace)$ is in general a finite function of the kinematical invariants.

The topology $C_4$ contains only soft singularities similar to those in $C_1$. The topologies $C_4\otimes C_4$ and
$C_4 \otimes C_1$ are, therefore, 
easier than $C_1\otimes C_1$. They can be treated exactly like $C_1\otimes C_1$ and we will not discuss them in what follows.

\subsection{Phase-space of  double real parton radiation}

We would like to point out that different parameterizations of the 
phase-space can factorize different sets of kinematic invariants. We will derive  two such 
parameterizations which allow  for   a  more  convenient numerical evaluation of diverse diagrams, according 
to their topology. 
 
\begin{figure}[htb!]
\centering%
\epsfig{file=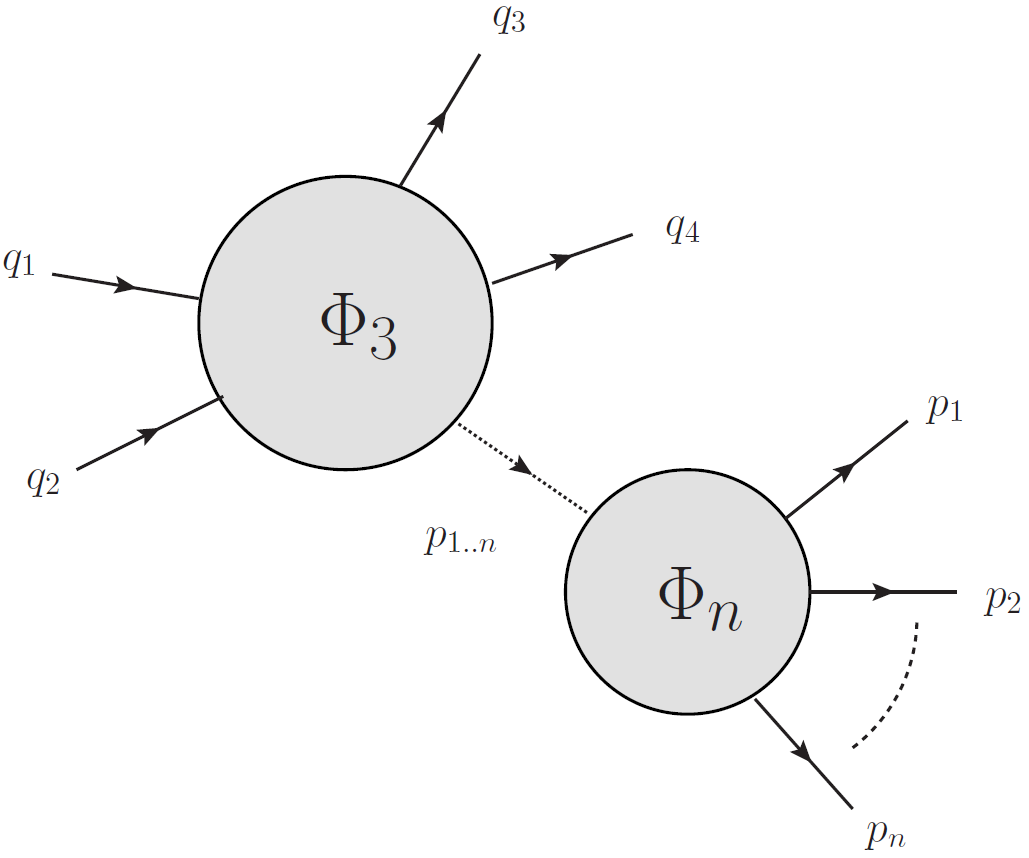,width=10cm}
\caption{Phase space factorization}
\label{fig:psfactorisation}
\end{figure}

The phase-space of $n$ massive particles in four dimensions is:
\begin{equation}
d\Phi_n(\sqrt{s};m_1,..,m_n)=(2\pi)^{4-3n}\left( \prod_{i=1}^n d^4p_i \delta^+(p_i^2-m_i^2)\right) \delta^{(4)}(q_1+q_2-\sum_{i=1}^n p_i), 
\end{equation}
where $s=(q_1+q_2)^2$. 
We  assume that a  $2\to n$ process exists at leading order in perturbation theory, and a strictly four-dimensional evaluation 
is therefore  sufficient. 
At NNLO, the double emission phase space is given by including two further massless particles (whose momenta we denote by $q_3$ and $q_4$)
\begin{eqnarray}
d\Phi_{n+2}(\sqrt{s};m_1,..,m_n,0,0) &=&
(2\pi)^{4-3n} 
\left( \prod_{i=1}^n d^4p_i \delta^+(p_i^2-m_i^2)\right) 
(2\pi)^{2-2d}
d^dq_3 \delta^+(q_3^2) d^dq_4 \delta^+(q_4^2) \nonumber \\
\times& &  \delta^{(d)}(q_1+q_2-\sum_{i=1}^{n} p_i-q_3-q_4).
\end{eqnarray}
We factorize the double real phase space into a 3-particle phase space times an n-particle phase space as follows
\begin{equation}
d\Phi_{n+2}(\sqrt{s};m_1,..,m_n,0,0)=\int \frac{ds_{1.. n}}{2\pi} d\Phi_3(\sqrt{s};0,0,\sqrt{s_{1.. n}}) d\Phi_n(\sqrt{s_{1.. n}};m_1,..,m_n).
\end{equation}
with $s_{1..n}=(\sum_{i=1}^n p_i)^2$ shall denote the center of mass energy (or invariant mass) of the n massive momenta $p_1,..,p_n$. This is depicted graphically  in Figure \ref{fig:psfactorisation}.
The limits of integration of $s_{1..n}$ are 
\begin{equation}
s \geq s_{1..n} \geq \left(\sum_{i=1}^n m_i\right)^2 
\end{equation}
and parameterizing $s_{1..n}$ linearly we get
\begin{equation}
s_{1..n}=\left(s-\left(\sum_{i=1}^n m_i\right)^2\right)x_5+\left(\sum_{i=1}^n m_i\right)^2.
\end{equation}
 
The parameter $x_5\in[0,1]$ then uniquely defines the double soft limit when $x_5=1$.
In the following discussion we will use the variable 
\begin{equation}
z=\frac{s_{1..n}}{s}
\end{equation}
which in the special case of $n=1$ reduces to $z=\frac{m_1^2}{s}$.
Then the variable $x_5$ is trivially removed and the double soft singularity occurs whenever $s=m_1^2$.

In the following we will assume that one can parametrize the $n$-particle phase space $d\Phi_n$, 
and we will focus on the phase-space of  the potentially unresolved  massless  partons $d\Phi_3$.

\subsection
	{
	Energies and angles parameterization
	}
\label{EAparm}

The three particle phase space element $d\Phi_3$ is
\beq
d\Phi_3(\sqrt{s};0,0,\sqrt{s_{1.. n}}) = (2\pi)^{3-2d}d^dq_3 \delta^{(+)}(q_3^2) d^dq_4 \delta^{(+)}(q_4^2) d^dQ \delta^{(+)}(Q^2-s_{1..n}) \delta^d(q_1+q_2-q_3-q_4-Q).
\eeq
Integrating out $Q$ and using that $d^dq\delta^{(+)}(q^2)=dE E^{d-3}d\Omega^{(d-1)}/2$ we get
\begin{eqnarray}
d\Phi_3(\sqrt{s};0,0,\sqrt{s_{1..n}}) &=& (2\pi)^{3-2d}\frac{1}{4} d\Omega_3^{(d-1)} d\Omega_4^{(d-1)}dE_3dE_4 (E_3 E_4)^{d-3} \nonumber \\
      \times  & &  \delta^{(+)}(s-s_{1..n}-2\sqrt{s}(E_3+E_4)+2E_3E_4(1-\cos\theta_{34})). 
\end{eqnarray}
We can solve the delta constraint for the energies in a symmetric way using the following ansatz:
\begin{eqnarray}
E_3  & = & \frac{1}{2}\sqrt{s}(1-z) x_1 \kappa \nonumber \\
E_4  & = & \frac{1}{2}\sqrt{s}(1-z) (1-x_1) \kappa. 
\end{eqnarray}
We find
\begin{equation}
\kappa=\frac{1-\sqrt{1-2(1-z)x_1(1-x_1)(1-\cos\theta_{34})}}{(1-z)x_1(1-x_1)(1-\cos\theta_{34})} \in [1,2) 
\end{equation}
The double soft limit now appears when $z\rightarrow1$, while the single soft singularities occur as $x_1\rightarrow0,1$.
After this transformation the phase space volume becomes
\begin{equation}
d\Phi_3=(2\pi)^{3-2d}\frac{d\Omega_3^{(d-1)} d\Omega_4^{(d-1)} dx_1 \kappa (1-z)}{16\sqrt{1-2x_1(1-x_1)(1-z)(1-\cos\theta_{34})}} \left(\frac{s(1-z)^2\kappa^2x_1(1-x_1)}{4}\right)^{d-3}.
\end{equation}

Having solved the energy constraint we move on to parametrize  the angles. 
Choosing the z-axis as the direction of $q_1$, we directly parameterize the angles which $q_3$ and $q_4$ make with the z-axis.
Finally we parametrize  the angle $\phi$ between $q_3$ and $q_4$ in the x-y plane 
leading to the following expressions of the solid angles
\begin{eqnarray}
d\Omega_3^{(d-1)} & = & d\Omega_3^{(d-2)} d\cos\theta_3 (\sin\theta_3)^{d-4} \nonumber \\
d\Omega_4^{(d-1)} & = & d\Omega_4^{(d-3)} d\cos\theta_4 (\sin\theta_4)^{d-4} d\cos\phi (\sin\phi)^{d-5}. 
\end{eqnarray}
Suppressing any extra dimensional components our 4-vectors are then fully parametrized as 
$q_3=E_3(1,\sin\theta_{3},0,\cos\theta_{3})$ and $q_4=E_4(1,\sin\theta_{4}\sin\phi,\sin\theta_{4}\cos\phi,\cos\theta_{4})$.
Mapping the remaining angles linearly, i.e. $\cos\theta_{3}=2x_3-1$, $\cos\theta_{4}=2x_4-1$ and $\phi=x_2\pi$, one obtains 
\begin{eqnarray}
\int d\Phi_3 & = & \frac{(2\pi)^{-3+2\epsilon}}{16\Gamma(1-2\epsilon)} \int_0^1dx_1dx_2dx_3dx_4    \left(\frac{s(1-z)^3\kappa^{4}x_1(1-x_1) }{2-\kappa}\right)  \nonumber \\
     \times  &   & \left(s^2 (1-z)^4 \kappa^4  x_1^2(1-x_1)^2x_3(1-x_3)x_4(1-x_4)\sin^2(\pi x_2) \right)^{-\epsilon}   
\end{eqnarray}
The following lists the propagators of  massless  partons in this parameterization:
\begin{eqnarray}
s_{13}  &=&-s(1-z)\kappa x_1 x_3      \nonumber \\  
s_{23}  &=&-s(1-z)\kappa x_1 (1-x_3) \nonumber \\
s_{14}  &=&-s(1-z)\kappa (1-x_1) x_4  \nonumber \\  
s_{24}  &=&-s(1-z)\kappa (1-x_1) (1-x_4) 
\end{eqnarray}
and
\begin{eqnarray}
s_{34}  & = & s(1-z)^2\kappa^2x_1(1-x_1)\tilde{x}_{34} \nonumber \\
s_{134} & = & s(1-z)\kappa\left[ (1-z)\kappa x_1(1-x_1)\tilde{x}_{34} - x_1 x_3  -(1-x_1) x_4\right] \nonumber \\
s_{234} & = & s(1-z)\kappa\left[ (1-z)\kappa x_1(1-x_1)\tilde{x}_{34} - x_1 (1-x_3)  -(1-x_1) (1-x_4))\right] 
\end{eqnarray}
where 
\begin{equation}
\tilde{x}_{34}=x_3+x_4-2x_3x_4-2\cos(x_2\pi)\sqrt{x_3(1-x_3)x_4(1-x_4)}
\end{equation}
and 
\begin{equation}
\kappa=\frac{1-\sqrt{1-4(1-z)x_1(1-x_1)\tilde{x}_{34}}}{2(1-z)x_1(1-x_1)\tilde{x}_{34}}.
\end{equation}

The angle between $q_3$ and $q_4$ is related to
\begin{equation}
\tilde{x}_{34}=\frac{1-\cos\theta_{34}}{2}=\frac{1-\cos\theta_3\cos\theta_4-\cos\phi\sin\theta_3\sin\theta_4}{2}. 
\end{equation}
This expression exposes the weak point of this parameterization:  it  gives rise to an overlapping line singularity when $\phi=0$ and $\theta_3=\theta_4$ i.e. when $q_3$ is parallel to $q_4$. 
Nevertheless the above construction can be used to fully subtract all phase space integrals which do not contain singularities in $\tilde{x}_{34}$, i.e. which do not contain $s_{34}$,$s_{134}$,$s_{234}$.

Let us now analyze the singularities in this parameterization. 
While $s_{13},s_{23},s_{14}$ and $s_{24}$ are fully factorized,
there is a overlapping line singularity in $s_{34}$ when $\tilde{x}_{34}=0$.\\
Furthermore there are overlapping singularities in $s_{134}$ and $s_{234}$. 
For $s_{134}$ there are $3$ different possibilities
\begin{eqnarray}
a) \, x_3=0 &\textrm{and}& x_4=0 \nonumber \\
b) \, x_3=0 &\textrm{and}& x_1=1 \nonumber \\
c) \, x_4=0 &\textrm{and}& x_1=0 
\end{eqnarray}
while for $s_{234}$ the singularities are located at
\begin{eqnarray}
a) \, x_3=1 &\textrm{and}& x_4=1  \nonumber \\
b) \, x_3=1 &\textrm{and}& x_1=1  \nonumber \\
c) \, x_4=1 &\textrm{and}& x_1=0. 
\end{eqnarray}
We can now apply this parameterization to all integrals of type $C_1 \otimes C_1$,$C_3 \otimes C_3$ 
and $C_1\otimes C_3$.

\subsubsection{
			Line singularities in the energy and angles parameterization
			}
\label{EAline}

One can use a non-linear transformation to get rid of the overlapping structure 
in $\tilde{x}_{34}$~\cite{Anastasiou:2003gr}.  A convenient way to derive such a mapping is remapping 
$\tilde{x}_{34}$ from $\tilde{x}_{34}^-=\tilde{x}_{34}(\phi=0)$ to $\tilde{x}_{34}^+=\tilde{x}_{34}(\phi=1)$ using
\begin{equation}
\tilde{x}_{34}=\frac{\tilde{x}_{34}^-\tilde{x}_{34}^+}{\tilde{x}_{34}^+ -x_2(\tilde{x}_{34}^+-\tilde{x}_{34}^-)}
\end{equation}
 
It is then apparent that $\tilde{x}_{34}$ will vanish whenever $\tilde{x}_{34}^-$ or $\tilde{x}_{34}^+$ will, for any value of $x_2$.\\
And that the overlapping line singularity is then re-casted into just a line singularity.
To aid numerical stability we perform the mapping $x_2\rightarrow (1-\cos(x_2\pi))/2$, such that 
$\tilde{x}_{34}$ becomes
\begin{equation}
\tilde{x}_{34}=\frac{(x_3-x_4)^2}{x_3+x_4-2x_3x_4+2\cos(x_2\pi)\sqrt{x_3(1-x_3)x_4(1-x_4)}}.
\end{equation} 
This is in fact identical to the mapping in~\cite{Czakon:2010td}.
The phase space volume then becomes
\begin{eqnarray}
\Phi_3 & = & \frac{(2\pi)^{-3+2\epsilon}}{16\Gamma(1-2\epsilon)} \int_0^1dx_1dx_2dx_3dx_4    \left(\frac{s(1-z)^3\kappa^{4}x_1(1-x_1) }{2-\kappa}\right)  \nonumber \\
      \times &   & \left(s^2 (1-z)^4 \kappa^4  x_1^2(1-x_1)^2x_3(1-x_3)x_4(1-x_4)\sin^2(\pi x_2) \right)^{-\epsilon}   \left(\frac{\tilde{x}_{34}}{|x_3-x_4|}\right)^{1-2\epsilon}.   
\end{eqnarray}
To factorize the line singularity in $s_{34}$ (at $x_3=x_4$) we are forced to split the integration region in two, separating
$x_3<x_4$  from $x_4<x_3$. 

\subsection
	{
	Hierarchical parameterization
	}
\label{Hparm}
Since in the energy and angles parameterization the invariants $s_{34},s_{134},s_{234}$ had line and overlapping singularities,
it is worth having a second parameterization which factorizes these, but may not factorize the others.
Our second parameterization closely resembles the features of the rapidity parameterization published in \cite{fehip}, 
however it is somewhat simpler. In this parameterization  
the three particle phase space element $d\Phi_3$ is
\beq
d\Phi_3(\sqrt{s};0,0,\sqrt{s_{1.. n}}) = (2\pi)^{3-2d}d^dq_3 \delta^{(+)}(q_3^2) d^dq_4 \delta^{(+)}(q_4^2) d^dQ \delta^{(+)}(Q^2-s_{1..n}) \delta^d(q_1+q_2-q_3-q_4-Q)
\eeq
is first factorized into 
a product of two 2-particle phase spaces
\begin{equation}
d\Phi_3(\sqrt{s},0,0,\sqrt{s_{1..n}})=\int \frac{ds_{34}} {2\pi} d\Phi_2(\sqrt{s},\sqrt{s_{34}},\sqrt{s_{1..n}}) d\Phi_2(\sqrt{s_{34}},0,0).
\end{equation}
with
\beq
d\Phi_2(\sqrt{s},\sqrt{s_{34}},\sqrt{s_{1..n}})= (2\pi)^{2-d} d^dQ \delta^{(+)}(Q^2-s_{1..n}) d^d\tilde{Q} \delta^{(+)}(\tilde{Q}^2-s_{34})\delta^d(q_1+q_2-\tilde{Q}-Q)
\eeq
and
\beq
d\Phi_2(\sqrt{s_{34}},0,0)= (2\pi)^{2-d}d^dq_3 \delta^{(+)}(q_3^2) d^dq_4 \delta^{(+)}(q_4^2)\delta^d(\tilde{Q}-q_3-q_4).
\eeq
We can parameterize $d\Phi_2(\sqrt{s},\sqrt{s_{34}},\sqrt{s_{1..n}})$ in terms of $s_{134}$, yielding
\beq
d\Phi_2(\sqrt{s},\sqrt{s_{34}},\sqrt{s_{1..n}})= (2\pi)^{2-d}\frac{1}{4s} d\Omega^{d-2} (\tilde{Q}_\perp)^{d-4} ds_{134}.
\eeq
To satisfy $\tilde{Q}_\perp\geq0$, we take
\begin{eqnarray}
0 &\leq& s_{34}  \leq \frac{s_{134} (s+s_{134}-s_{1..n})}{s_{134}-s_{1..n}} \nonumber \\
0 &\geq& s_{134} \geq (s_{1..n}-s). 
\end{eqnarray}
$d\Phi_2(\sqrt{s_{34}};0,0)$ can be parameterized in terms of the invariants $s_{13}$ and $s_{23}$ yielding
\beq
d\Phi_2(\sqrt{s_{34}};0,0)=(2\pi)^{2-d}\frac{1}{8\tilde{Q}_\perp s} ds_{13}ds_{23} d\Omega^{d-3}  \left[(p_3)_\perp \sin\phi\right]^{d-5}
\eeq
where $\phi$ is the angle between $(p_3)_\perp$ and $\tilde{Q}_\perp$.
We fulfil the constraint $(p_3)_\perp \sin\phi \geq 0$ to find the limits of integration for $s_{13}$ and then for $s_{23}$.  \\
Parameterizing $s_{134},s_{34},s_{13}$ and $s_{23}$ linearly we arrive at
\begin{eqnarray}
\int d\Phi_3 & = & \frac{(2\pi)^{-3+2\epsilon}}{16\Gamma(1-2\epsilon)} \int_0^1 dx_1dx_2dx_3dx_4    \left(\frac{s(1-z)^3x_1(1-x_1)}{z+x_1(1-z)}\right)  \nonumber \\
  \times     &   & \left(\frac{s^2 (1-z)^4  x_1^2(1-x_1)^2x_2(1-x_2)x_3(1-x_3)\sin^2(\pi x_4)}{z+x_1(1-z)} \right)^{-\epsilon}.   
\end{eqnarray}
The invariants in this parameterization are 
\begin{eqnarray}
s_{34}  & = & \frac{s(1-z)^2 x_1(1-x_1)x_2}{z+x_1(1-z)} \nonumber \\
s_{134} & = & -s(1-z)x_1 \nonumber \\
s_{234} & = & -s(1-z)(1-x_1)\left[\frac{z+x_1(1-x_2)(1-z)}{z+x_1(1-z)} \right]  \nonumber \\
s_{23}  & = & -s(1-z)(1-x_1)x_3 \nonumber \\
s_{24}  & = & -s(1-z)(1-x_1)(1-x_3) 
\end{eqnarray}
and
\begin{eqnarray}
s_{13}  &=& -s(1-z)x_1  \bigg[ x_3(1-x_2)   
          +\frac{x_2(1-x_3)}{z+x_1(1-z)}-2\cos(\pi x_4) \sqrt{\frac{x_2(1-x_2)x_3(1-x_3)}{z+x_1(1-z)}} \bigg]       \nonumber \\    
s_{14}  &=& -s(1-z)x_1  \bigg[ (1-x_3)(1-x_2)      
           +\frac{x_2x_3}{z+x_1(1-z)}+2\cos(\pi x_4) \sqrt{\frac{x_2(1-x_2)x_3(1-x_3)}{z+x_1(1-z)}} \bigg]. \nonumber         
\end{eqnarray}
We see that the only invariants which are not factorized are $s_{13}$ and $s_{14}$. The variable
$s_{13}$ contains overlapping singularities at $x_3=0=x_2$ and $x_3=1=x_2$ as well as 
an overlapping line singularity at $x_4=0,x_1=1,x_3=x_2$, while
$s_{14}$ contains overlapping singularities at $x_3=0,x_2=1$ and $x_3=1,x_2=1$ 
as well as an overlapping line singularity at $x_4=1,x_1=1,x_3=1-x_2$.

\subsubsection{Line singularities in the hierarchical parameterization}
\label{Hresolve}

Consider the expressions
\begin{equation}
\frac{J(p_1,p_2,p_3,p_4)}{s_{13}s_{24}}\;\;\; , \;\;\;\frac{J(p_1,p_2,p_3,p_4)}{s_{13}s_{23}}
\end{equation}
with $J(p_1,p_2,p_3,p_4)$ a finite numerator function.
They both contain a line singularity due to  $s_{13}$ in the denominator. We now use the  partial fractioning identities, 
\begin{equation}
\frac{1}{s_{13}s_{24}}=\frac{1}{s_{13}s_{234}+s_{134}s_{24}}\left( \frac{s_{134}}{s_{13}} + \frac{s_{234}}{s_{24}} \right), 
\end{equation}
\begin{equation}
\frac{1}{s_{13}s_{23}}=\frac{1}{s_{13}s_{234}+s_{23}s_{134}}\left( \frac{s_{134}}{s_{13}} + \frac{s_{234}}{s_{23}} \right).
\end{equation}
The term $s_{13}s_{234}+s_{134}s_{24}$ has an overlapping singularity at $x_3=1=x_2$,
while the term $s_{13}s_{234}+s_{134}s_{23}$ has an overlapping singularity at $x_3=0=x_2$. 
Then we exchange $1\leftrightarrow2$ and $3\leftrightarrow4$ in the term containing $s_{13}$ to rotate the line singularity out,
i.e.
\begin{equation}
\frac{J(p_1,p_2,p_3,p_4)}{s_{13}s_{24}}=\frac{J(p_1,p_2,p_3,p_4)+J(p_2,p_1,p_4,p_3)}{s_{13}s_{234}+s_{134}s_{24}}\frac{s_{234}}{s_{24}} 
\end{equation}
\begin{equation}
\frac{J(p_1,p_2,p_3,p_4)}{s_{13}s_{23}}=\frac{J(p_1,p_2,p_3,p_4)+J(p_2,p_1,p_3,p_4)}{s_{13}s_{234}+s_{23}s_{134}} \frac{s_{234}}{s_{23}}.
\end{equation}
and we are left with just overlapping singularities, which can be treated as explained in the following section. 
This trick was first discovered by Frank Petriello~\cite{FrankPrivate} and it has been used in the 
implementation of the program \fehip   ~described in~\cite{fehip}, it has been also been used in the evaluation of doublereal counterterms
in~\cite{gabor}.


\section{
		Numerical evaluation of double-real radiation phase-space integrals
		}
\label{sec:RRnumerics}

In this section, we present a numerical evaluation of all types of 
scalar phase-space integrals  which appear in NNLO calculations. To evaluate our integrals numerically 
we choose the point $(s=1,z=0.1)$.  We will use the notation 
\begin{equation}
\bar{x}_i=1-x_i,  
\end{equation}
where the $x_{i}\in [0,1]$ are parameters of integration.
\begin{enumerate}
  \item Topology $C_1 \otimes C_1$: 
      \begin{enumerate}
       \item
	    The integral
	    \begin{equation}
	    I_{11a}= \int \frac{d\Phi_3}{s_{13}s_{23}s_{14}s_{24}}
	    \end{equation}
	    fully factorizes in the energies and angles parameterization (Section \ref{EAparm}), we obtain
	    \begin{equation}
	    I_{11a}= 0.09400(2)+\frac{0.010951(4)}{\epsilon}-\frac{0.0035586(5)}{\epsilon^2}-\frac{0.001119844946(1)}{\epsilon^3}.
	    \end{equation}
      \item
	    The integral
	    \begin{equation}
	    I_{11b}=\int d\Phi_3 \frac{(s_{34}s-s_{14}s_{23})^2}{s_{13}^2s_{24}^2}
	    \end{equation}
	    with the numerator structure as in \cite{fehip},
	    factorizes in the energies and angles parameterization (Section \ref{EAparm}), we get      
	    \begin{equation}
	    I_{11b}=0.023885(3)+\frac{0.0041606(3)}{\epsilon}+\frac{0.00036930(4)}{\epsilon^2}. 
	    \end{equation}
      \end{enumerate}

      \item Topology $C_2 \otimes C_2$:\\
	\begin{enumerate}
	  \item   
	      The integral
	      \begin{equation}
	      I_{22a}=\int \frac{d\Phi_3 N(\left\lbrace s_{ij}\right\rbrace)}{(s_{34}s_{134})^2} 
	      \end{equation}
	      factorizes in the hierarchical parameterization (Section \ref{Hparm}). 
	      The numerator function has the scaling behavior $N(\left\lbrace s_{ij}\right\rbrace)\sim s_{34}s_{134}$ \cite{fehip}.
	      We obtain
	      \begin{equation}
	      I_{22a}=\int \frac{d\Phi_3}{s_{34}s_{134}}=0.0011728(1)-\frac{0.00050726(1)}{\epsilon}-\frac{0.000125982556(0)}{\epsilon^2}. 
	      \end{equation}	
	  \item 		  
	      The integral
	      \begin{equation}
	      I_{22b}=\int \frac{d\Phi_3 N(\left\lbrace s_{ij}\right\rbrace)}{s_{34}^2s_{134}s_{234}} 
	      \end{equation}
	      factorizes in the hierarchical parameterization (Section \ref{Hparm}). The numerator 
	      scales as $N(\left\lbrace s_{ij}\right\rbrace)\sim s_{34}$. We obtain
	      \begin{equation}
	      I_{22b}=\int \frac{d\Phi_3 }{s_{34}s_{134}s_{234}} =-0.0015003(2)+\frac{0.00112726(5)}{\epsilon}+\frac{0.000279961236(1)}{\epsilon^2}.
	      \end{equation}
	\end{enumerate}

  \item Topology $C_3 \otimes C_3$:
      \begin{enumerate}
       \item 
	  The integral
	  \begin{equation}
	  I_{33a}=\int d\Phi_3 \frac{(s_{34}s-s_{14}s_{23})^2}{s_{234}^2s_{24}^2} 
	  \end{equation}
	  factorizes in the hierarchical parameterization (section \ref{Hparm}). The numerator structure can be found in~\cite{fehip}.  We obtain
	  \begin{equation}
	  I_{33a}=-0.003841(2)+\frac{0.0007814(4)}{\epsilon}+\frac{0.00018465(1)}{\epsilon^2}. 
	  \end{equation}
      \item
	  The integral
	  \begin{equation}
	  I_{33b}=\int  \frac{d\Phi_3}{s_{134}s_{234}s_{13}s_{23}}
	  \end{equation}
	  neither factorizes in energies and angles nor in the hierarchical parameterization. 
	  We use the hierarchical parameterization (section \ref{Hparm}), since fewer overlapping singularities are present there.
	  Using partial fractions, as described in section~\ref{Hresolve} we can rewrite the integral as 
	  \begin{equation}
	  I_{33b}=\int  \frac{2d\Phi_3 }{s_{23}s_{134}(s_{134}s_{23}+s_{234}s_{13})}.
	  \end{equation}
	  This contains the following substructure
	  \begin{equation}
	  \frac{1}{x_3} \frac{1}{x_3A+ x_2\bar{x}_3 B+C\sqrt{x_2\bar{x}_2x_3\bar{x}_2}}
	  \end{equation}
	  with $A,B,C$ finite. This becomes singular when $x_3=0=x_2$ where $x_3$ is active. 
	  We factorize this singularity by applying
  	  \begin{equation}
	  x_3 \rightarrow \frac{x_3x_2}{(1-x_3)+x_2}	   
	  \end{equation}
	  and obtain
	  \begin{equation}
	  I_{33b}=0.023155(3)+ \frac{0.0076371(1)}{\epsilon}+ \frac{0.00007730(6)}{\epsilon^2}-\frac{0.000279961236(1)}{\epsilon^3}.
	  \end{equation}
      \item
	  The integral
	  \begin{equation}
	  I_{33c}=\int  \frac{d\Phi_3}{s_{134}s_{234}s_{13}s_{24}}
	  \end{equation}
	  is similar to $I_{33b}$ in the hierarchical parameterization (Section \ref{Hparm}). Partial fractioning as before we get
	  \begin{equation}
	  I_{33c}=\int  \frac{2d\Phi_3 }{s_{24}s_{134}(s_{134}s_{24}+s_{234}s_{13})}.
	  \end{equation}
	  This contains the substructure
	  \begin{equation}
	  \frac{1}{\bar{x}_3} \frac{1}{ \bar{x}_2 A + \bar{x}_3 x_2 B+C\sqrt{x_2\bar{x}_2x_3\bar{x}_3}}
	  \end{equation}
	  with $A,B,C$ finite. This
	  becomes singular when $\bar{x}_3=0=\bar{x}_2$ with $\bar{x}_3$ being active. 
	  We disentangle this singularity by applying
	  \begin{equation}
	  \bar{x}_3\rightarrow \frac{\bar{x}_3\bar{x}_2}{(1-\bar{x}_3)+\bar{x}_2}.	   
	  \end{equation}
	  We then obtain
	  \begin{equation}
	  I_{33c}=0.12567(9) -\frac{0.03645(1)}{\epsilon} -\frac{0.018566(1)}{\epsilon^2} +\frac{0.002799612364(0)}{\epsilon^3}. 
	  \end{equation}
    \end{enumerate}

 \item Topology $C_1\otimes C_3$:\\
    \begin{enumerate}
       \item   
	  The integral
	  \begin{equation}
	  I_{13a}=\int d\Phi_3 \frac{(s_{34}s-s_{14}s_{23})^2}{s_{234}s_{24}^2s_{13}} 
	  \end{equation}
	  with the numerator structure as in~\cite{fehip}. The singularities
	  factorize in energies and angles. We immediately obtain
	  \begin{equation}
	  I_{13a}=-0.0040885(4)-\frac{0.00036930(1)}{\epsilon}. 
	  \end{equation}
       \item   	  
	  The integral
	  \begin{equation}
	  I_{13b}=\int d\Phi_3 \frac{N(\left\lbrace s_{ij}\right\rbrace)}{s_{134}s_{13}s_{14}s_{23}}  
	  \end{equation}   
	  has a quadratic divergence due to the term $s_{134}s_{13}s_{14}$. 
          This means that $N(\left\lbrace s_{ij}\right\rbrace)\sim \left \lbrace s_{134},s_{13},s_{14} \right \rbrace $.
	  Such that
	  \begin{equation}
	  I_{13b}=\left \lbrace \int  \frac{d\Phi_3}{s_{13}s_{14}s_{23}},\int  \frac{d\Phi_3}{s_{134}s_{14}s_{23}} \right \rbrace   
	  \end{equation}
	  the first of which is a sub-topology of $C_1^2$ while the second is a sub-topology of $C_3^2$.
    \end{enumerate}

 \item Topology $C_1\otimes C_2$:
  \begin{enumerate}
    \item  
    The integral  
    \begin{equation}
    I_{12}=\int \frac{d\Phi_3 N(\left\lbrace s_{ij}\right\rbrace)}{s_{34}s_{234}s_{13}s_{24}}   
    \end{equation}
    has a quadratic divergence due to the term $s_{34}s_{234}s_{24}$.
    The numerator can have the following scalings: $N(\left\lbrace s_{ij}\right\rbrace)\sim \left \lbrace s_{34},s_{234},s_{24} \right \rbrace $.
    We therefore consider the following possibilities
    \begin{equation}
    I_{12}=\left \lbrace \int\frac{d\Phi_3}{s_{34}s_{13}s_{24}}
			,\int\frac{d\Phi_3}{s_{34}s_{234}s_{13}}			 
			,\int\frac{d\Phi_3}{s_{234}s_{13}s_{24}}.  \right \rbrace   
    \end{equation}
    The last of these is a sub-topology of $C_3^2$ and does not merit further attention. 	
    We will evaluate the other two in the hierarchical parameterization.
    For
    \begin{equation}
    I_{12a}= \int \frac{d\Phi_3}{s_{34}s_{13}s_{24}}
    \end{equation}
    we use the same strategy as we used for $I_{33c}$. We obtain
    \begin{equation}
    I_{12a} =   0.07115(1)+\frac{0.006996(1)}{\epsilon}-\frac{0.0029912(1)}{\epsilon^2}-\frac{0.000839883709(0)}{\epsilon^3}.
    \end{equation}
    The integral
    \begin{equation}
    I_{12b}=\int \frac{d\Phi_3}{s_{34}s_{234}s_{13}}
    \end{equation}
    factorizes in the hierarchical parameterization. We obtain
    \begin{equation}
    I_{12b} =   0.0198554(9)+\frac{0.0023667(2)}{\epsilon}-\frac{0.00088965(4)}{\epsilon^2}-\frac{0.000279961236(1)}{\epsilon^3}.
    \end{equation}

  \end{enumerate}

 \item Topology $C_2\otimes C_3$:
  \begin{enumerate}
    \item
      The integral
      \begin{equation}
       I_{23a}=\int \frac{d\Phi_3N(\left\lbrace s_{ij}\right\rbrace)}{s_{34}s_{234}^2s_{23}}
      \end{equation}	
      factorizes in the hierarchical parameterization, but carries a cubic divergence in $(1-x_1)\sim s_{234}$. 
      Taking $N(\left\lbrace s_{ij}\right\rbrace)\sim s_{234}^2$, we get
      \begin{equation}
       I_{23a}=\int \frac{d\Phi_3 }{s_{34}s_{23}}
      \end{equation}
      which just is a sub-topology of $I_{12a}$. While other numerators are possible
      these do not give different singularity structures.
      
    \item 
      The integral   
      \begin{equation}
       I_{23a}=\int \frac{d\Phi_3N(\left\lbrace s_{ij}\right\rbrace)}{s_{34}s_{134}s_{234}s_{23}}
      \end{equation}	
       factorizes in the hierarchical parameterization, but carries a quadratic divergence in $(1-x_1)\sim s_{234}$. 
       A minimal choice for the numerator is $N(\left\lbrace s_{ij}\right\rbrace) \sim s_{234}$ in which case we recover $I_{22b}$.
       Hence no new singularity structures can be obtained from this topology.  
  \end{enumerate}

 \item 
  We will now consider interferences of $C_4$ with $C_2$ and $C_3$. 
  One can evaluate these interferences in the energies and angles parameterization. 
  In the following we will use  $t_{13}\sim E_{3} \sim (s_{13}+s_{23})$ 
  and $t_{24}\sim E_{4} \sim (s_{14}+s_{24})$.
  \begin{enumerate}
    \item Topology $C_2\otimes C_4$:\\
      The integral
      \begin{equation}
       I_{24}=\int \frac{d\Phi_3}{s_{34}s_{134}(s_{13}+s_{23})(s_{14}+s_{24})}
      \end{equation}
      has the following singularity structure
      \begin{eqnarray}
      \frac{1}{x_1 \bar{x}_1 (x_3-x_4)}  \frac{1}{A x_1 \bar{x}_1 (x_3-x_4)^2 +B x_1 x_3 + C \bar{x}_1 x_4} 
      \end{eqnarray}
      in the energy and angle parameterization after the mapping (see \ref{EAline}) is applied.
      We first split the integration region into two sectors which we define as $x_3<x_4$ (sector 1) and $x_4<x_3$ (sector 2). 
      After this sector decomposition we are still left with overlapping singularities at $x_3=0=\bar{x}_1$ in sector 1
      and at $x_4=0=x_1$ in sector 2.
      These can be disentangled using 
      \begin{equation}
      \bar{x}_1\rightarrow \frac{\bar{x}_1x_3}{(1-\bar{x}_1)+x_3}
      \end{equation}
      in sector 1 and
      \begin{equation}
      x_1\rightarrow \frac{x_1 x_4}{(1-x_1)+x_4}
      \end{equation}
      in sector 2.
      We then obtain
      \begin{equation}
       I_{24}=-0.006956(3)-\frac{0.0010708(3)}{\epsilon}+\frac{0.00065900(1)}{\epsilon^2}+\frac{0.000207378694(0)}{\epsilon^3}.
      \end{equation}

    \item Topology $C_3\otimes C_4$:\\
      The integral
      \begin{equation}
       I_{34}=\int \frac{d\Phi_3 }{s_{13}s_{134}(s_{13}+s_{23})(s_{14}+s_{24})}
      \end{equation}
      has the following singularity structure
      \begin{eqnarray}
      \frac{1}{x_1 x_3}  \frac{1}{A x_1 \bar{x}_1 (x_3-x_4)^2 +B x_1 x_3 + C \bar{x}_1 x_4}. 
      \end{eqnarray}
      It contains no line singularity but several overlapping ones located at $x_3=0=x_4$, $x_4=0=x_1$ and at $x_3=0,\bar{x}_1=0$.
      To separate the two singularities we first partial fraction the soft singularities by multiplying by 
      $1=x_1+\bar{x}_1$. We then treat the two terms with different nonlinear transformations.
      For the first term we apply the mapping	
      \begin{equation}
      x_3\rightarrow \frac{x_3x_4\bar{x}_1}{(1-x_3)+x_4\bar{x}_1}.
      \end{equation}
      since $x_3$ is the only active singularity, it is clear that we had to remap it.
      The second term is more difficult, since both $x_3$ and $x_1$ are now active.
      We apply the following sequence of mappings:\\
      First let
      \begin{equation}
      x_3\rightarrow \frac{x_3 x_4}{(1-x_3)+x_4}
      \end{equation}
      and then 
      \begin{eqnarray}
       x_1 &\rightarrow& \frac{x_1 \bar{x}_3}{(1-x_1)+\bar{x}_3} \nonumber \\ 
       x_4 &\rightarrow& \frac{x_4 \bar{x}_3}{(1-x_4)+\bar{x}_3}. \\
      \end{eqnarray}
      We obtain
      \begin{equation}
       I_{34}=    -0.32519(4)-\frac{0.048942(2)}{\epsilon}-\frac{0.0062917(3)}{\epsilon^2}-\frac{0.000559922473(3)}{\epsilon^3}.
      \end{equation}

  \end{enumerate}

\end{enumerate}


\section{
		Two loop examples
		}
\label{sec:loops}
In what follows we show how one can use non-linear mappings to disentangle singularities in two-loop integrals appearing in NNLO virtual amplitudes. We treat only two indicative cases, the massless non-planar triangle with one leg off-shell and the massless non-planar box with all legs on-shell, due to their particularly intricate singularity structure. Integrals involving masses are in general simpler as far as  factorization of singularities is concerned. 

\subsection{
			The massless non-planar triangle with one leg off-shell.
		}
\begin{figure}[htb!]
\centering%
\epsfig{file=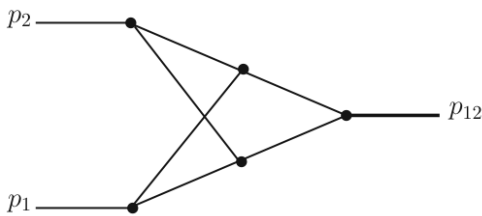,width=7cm}
\caption{The massless non-planar two-loop triangle with one legs off-shell}
\label{fig:cross_tri}
\end{figure}

The two-loop, non-planar triangle with one off-shell leg (see fig.~\ref{fig:cross_tri}) and momenta $p_1,p_2$ 
\beq
Xtri = \int \frac{d^dk_1}{i\pi^{d/2}} \frac{d^dk_2}{i\pi^{d/2}} \frac{1}{k_1^2 (k_1+p_1)^2 k_2^2 (k_2+p_2)^2 (k_1+k_2)^2 (k_1+k_2+p_1+p_2)^2}
\eeq
A  Feynman parameterization reads: 
\begin{equation}
Xtri = 4^{2+2\eps} 
\int_0^1 dx_1 dx_2 dz dy dx \frac{ z y^{1+\eps} (1-y)^{-1-\eps}  (1-z)^{-1-\eps} }{\left[  
x (1-x) + y z (x-x_1) (x-x_2)
\right]^{2 +2 \eps} }
\label{Xtri_initial}
\end{equation}

The  first overlapping  singularity is  at  $x=0$ or  $x=1$ and  $y=0$. We also notice that there is a  singularity at  $y=1$.  To avoid infinite  looping we must first guarantee, 
as in sector  decomposition, that  no singularities  occur  at the upper limit of integration.  

We  split $x$ in the two intervals $R_a = [0, 1/2]$ and  $R_b = [1/2, 1]$  and map the integration region back to the unit hypercube.  
In $R_a$, 
\[ 
x \to x/2,
\]
and in $R_b$, 
\[ 
x \to 1 -x/2, \quad x_1 \to 1-x_1, \quad x_2 \to 1-x_2.
\]
This gives  two identical integrals (the integral in eq.\ref{Xtri_initial} is invariant  under the combined $x\rightarrow 1-x , \; x_1\rightarrow 1-x_1,\; x_2\rightarrow 1-x_2$), and we can write
\begin{equation}
Xtri = 4^{2+2\eps} 
\int_0^1 dx_1 dx_2 dz dy dx \frac{ z y^{1+\eps} (1-y)^{-1-\eps}  (1-z)^{-1-\eps} }{\left[  
x (2-x) + y z (2x_1 -x) (2x_2-x)
\right]^{2 +2 \eps} }
\label{Xtri_after_x_split}
\end{equation}

Note here that the denominator of eq.~\ref{Xtri_after_x_split} has the same singularity structure as 
\beq
x+yzx_1x_2 - yzx(x_1+x_2)
\eeq

In particular, there are still singularities at upper corners of the hypercube, when $y\rightarrow 1$ and $z\rightarrow 1$. This leads us to  split the $y$ integration region, $y \to y/2$ and  $y \to 1 -y/2$ and the $z \to z/2$ and $z\to 1-z/2$.

We obtain 
\begin{equation}
Xtri = Xtri_{11} + Xtri_{12} + Xtri_{21}+ Xtri_{22},
\end{equation}
where 
\begin{equation}
Xtri_{11} = 2^{6+9\eps}
\int_0^1 dx_1 dx_2 dz dy dx \frac{ z y^{1+\eps} (2-y)^{-1-\eps}  (2-z)^{-1-\eps} }{\left[  
4 x (2-x) + y z (2x_1 -x) (2x_2-x)
\right]^{2 +2 \eps} }
\end{equation}
\begin{equation}
Xtri_{12} = 2^{6+9\eps}
\int_0^1 dx_1 dx_2 dz dy dx \frac{ (2-z) y^{1+\eps} (2-y)^{-1-\eps}  z^{-1-\eps} }{\left[  
4 x (2-x) + y (2-z) (2x_1 -x) (2x_2-x)
\right]^{2 +2 \eps} }
\end{equation}
\begin{equation}
Xtri_{21} = 2^{6+9\eps}
\int_0^1 dx_1 dx_2 dz dy dx \frac{ z (2-y)^{1+\eps} y^{-1-\eps}  (2-z)^{-1-\eps} }{\left[  
4 x (2-x) + (2-y) z (2x_1 -x) (2x_2-x)
\right]^{2 +2 \eps} }
\end{equation}
\begin{equation}
Xtri_{22} =  2^{6+9\eps}
\int_0^1 dx_1 dx_2 dz dy dx \frac{ (2-z) (2-y)^{1+\eps} y^{-1-\eps}  z^{-1-\eps} }{\left[  
4 x (2-x) + (2-y) (2-z) (2x_1 -x) (2x_2-x)
\right]^{2 +2 \eps} }
\end{equation}

The first three sectors, $Xtri_{11},Xtri_{12},Xtri_{21}$ are free of singularities at 1. They can be directly treated by a non-linear mapping. 

$Xtri_{11}$ has a singularity structure equivalent to that of $x+yzx_1x_2$, similar to eq.~\ref{example2} and we use the mapping (directly analogous to eq.~\ref{mapping2})
\beq
\NLmap{x}{yz x_1 x_2}.
\eeq

$Xtri_{12}$ has a singularity structure equivalent to that of $x+yx_1x_2$, also similar to  eq.~\ref{example2} and we use the mapping
\beq
\NLmap{x}{y x_1 x_2}.
\eeq

$Xtri_{21}$ has a singularity structure equivalent to that of  $x+zx_1x_2$, also similar to  eq.~\ref{example2} and we use the mapping
\beq
\NLmap{x}{z x_1 x_2}.
\eeq

$Xtri_{22}$ is a bit more complicated. Its singularity structure is the one of 
\beq
x+x_1x_2 - x(x_1+x_2)
\eeq
 It retains singularities at $x_{1,2}\to 1$. We therefore split this integral further in $x_1 \to x_1/2$ and $x_1\to 1-x_1/2$ as well as $x2 \to x2/2$ and $x_2 \to 1-x_2/2$.

We obtain 
\begin{equation}
Xtri_{2211} =  2^{5+9\eps}
\int_0^1 dx_1 dx_2 dz dy dx \frac{ (2-z) (2-y)^{1+\eps} y^{-1-\eps}  z^{-1-\eps} }{\left[  
4 x (2-x) + (2-y) (2-z) (x_1 -x) (x_2-x)
\right]^{2 +2 \eps} }
\end{equation}

with a singularity structure equivalent to $x+x_1x_2$ (i.e. eq.~\ref{example2}) for which we will use the mapping 
\beq
\NLmap{x}{x_1x_2}.
\eeq

\begin{equation}
Xtri_{2212} =  2^{5+9\eps}
\int_0^1 dx_1 dx_2 dz dy dx \frac{ (2-z) (2-y)^{1+\eps} y^{-1-\eps}  z^{-1-\eps} }{\left[  
4 x (2-x) + (2-y) (2-z) (x_1 -x) (2-x_2-x)
\right]^{2 +2 \eps} }
\end{equation}

with a singularity structure equivalent to $x_1+x(x_1+y+z+x_2)$, similar to eq.~\ref{example4} for which we will use the following sequence of  mappings 
\beq
\NLmap{x_1}{x}
 \;\;,\;\;
 \NLmap{y}{x_2}
\;\;,\;\;
\NLmap{z}{x_1}
 \;\;,\;\;
 \NLmap{x_1}{x_2}
\eeq

\begin{equation}
Xtri_{2221} =  2^{5+9\eps}
\int_0^1 dx_1 dx_2 dz dy dx \frac{ (2-z) (2-y)^{1+\eps} y^{-1-\eps}  z^{-1-\eps} }{\left[  
4 x (2-x) + (2-y) (2-z) (2-x_1 -x) (x_2-x)
\right]^{2 +2 \eps} }
\end{equation}
with a singularity structure equivalent to $x_2+x(x_2+y+z+x_1)$,similar to eq.~\ref{example4},  for which we will use the following sequence of  mappings 
\bea
\NLmap{x_2}{x}\;\;,\;\;
\NLmap{y}{x_1}y \;\;,\;\;
\NLmap{z}{x_2}
\;\;,\;\;
\NLmap{x_2}{x_1}
\eea

\begin{equation}
Xtri_{2222} =  2^{5+9\eps}
\int_0^1 dx_1 dx_2 dz dy dx \frac{ (2-z) (2-y)^{1+\eps} y^{-1-\eps}  z^{-1-\eps} }{\left[  
4 x (2-x) + (2-y) (2-z) (2-x_1 -x) (2-x_2-x)
\right]^{2 +2 \eps} }
\end{equation}
which is finite!

We therefore end up with 7 different integrals to be numerically evaluated.  This should be contrasted with the 64 number of sectors one arrives using sector decomposition. The numerical convergence of these integrals poses no additional problems and we have checked that the numerical result agrees with the analytic result known in the literature.

\subsection{
			The non-planar double box
		}

\begin{figure}[htb!]
\centering%
\epsfig{file=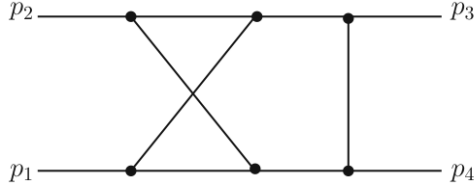,width=7cm}
\caption{The massless non-planar double box with all legs on-shell}
\label{fig:cross_box}
\end{figure}

Using the representation of ref.~\cite{AnastasiouBanfi} for the two loop non-planar box (see fig.~\ref{fig:cross_box}), we get the expression
\beq
{\rm Xbox} = C_{\epsilon} 
		\int{
			\frac{dx_1dx_2dx_3dx_4\delta(1-x_1-x_2-x_3-x_4) d\tau_1d\tau_2\,\,x_2^{1+\epsilon}}
				{ (x_1x_3s + x_2x_4 t_c + x_1x_2Q^2 + x_2x_3Q_{t}^2 
					)^{3+2\epsilon}}
					}
\eeq

where
\beq
Q_t^2= (1-\tau_1)(1-\tau_2)s,\;\;\;\;
Q^2= \tau_1\tau_2s,\;\;\;\;
t_c=\tau_2(1-\tau_1)u+(1-\tau_2)\tau_1t
\eeq
and
\beq
C_\epsilon=\frac{2\Gamma(3+2\epsilon) \Gamma(-\epsilon) \Gamma(1-\epsilon)}{\Gamma(1+\epsilon)^2 \Gamma(1-2\epsilon)}
\eeq

In order to avoid the singularities at the upper corners of the integration region we split the integral in four, mapping $\tau_1 \to \tau_1/2$, $\tau_1 \to 1-\tau_1/2$ and then $\tau_2 \to \tau_2/2$ and $\tau_2 \to 1-\tau_2/2$. Two of the resulting integrals can be mapped to the other two by exchanging $x1$ and $x_3$, so we end up with
\beq
{\rm Xbox}_a =C_{\epsilon}  \frac{4^{3+2\epsilon}}{2}
		\int{
			\frac{dx_1dx_2dx_3dx_4\delta(1-x_1-x_2-x_3-x_4) d\tau_1d\tau_2\,\,x_2^{1+\epsilon}}
				{ (4x_1x_3s + x_1x_2 B_1B_2s+ x_2x_3\tau_1\tau_2s 
					+ x_2x_4 ( \tau_2B_1t+B_2\tau_1u)  )^{3+2\epsilon}}
					}
\eeq

\beq
{\rm Xbox}_b =C_{\epsilon}  \frac{4^{3+2\epsilon}}{2}
		\int{
			\frac{dx_1dx_2dx_3dx_4\delta(1-x_1-x_2-x_3-x_4) d\tau_1d\tau_2\,\,x_2^{1+\epsilon}}
				{ (4x_1x_3s + x_1x_2 \tau_1B_2 s+ x_2x_3B_1\tau_2s 
					+ x_2x_4 ( B_1B_2u+\tau_1\tau_2t)  )^{3+2\epsilon}}
					}
\eeq
where
\beq
B_{1,2}\equiv 2-\tau_{1,2}
\eeq
Subsequently, we use the method of primary sectors on the variables $x_1,x_2,x_3,x_4$ on each of the above integrals, to get eight primary sectors:
\begin{enumerate}
\item
${\rm Xbox}_{a1}$:  ${\rm Xbox}_a$ where $x_1>x_{2,3,4}$ 
has the singularity structure of $x_2+x_3$, and we use the mapping of eq.~\ref{eq:transform}.
\item
${\rm Xbox}_{a2}$: ${\rm Xbox}_a$ where $x_2>x_{1,3,4}$ 
has the rather intricate singularity structure of the example eq.~\ref{example_sd_4}. We follow the discussion given there and  decompose it to six integrals.

\item
${\rm Xbox}_{a3}$:  ${\rm Xbox}_a$ where $x_3>x_{1,2,4}$ 
has the rather intricate singularity structure of the example eq.~\ref{example_sd_3}. We follow the discussion given there and  decompose it to five integrals.

\item
${\rm Xbox}_{a4}$:  ${\rm Xbox}_a$ where $x_4>x_{1,2,3}$ 
has the singularity structure of eq.~\ref{example_sd_2}. We follow the discussion given there and  decompose it in two sub-sectors, in $x_2,x_3$, each of which can be factorized.

\item
${\rm Xbox}_{b1}$:  ${\rm Xbox}_b$ where $x_1>x_{2,3,4}$ 
has the singularity structure of eq.~\ref{example4} and we use the mapping of eq.~\ref{mapping4}.

\item
${\rm Xbox}_{b2}$:  ${\rm Xbox}_b$ where $x_2>x_{1,3,4}$ 
has the singularity structure of eq.~\ref{example_xsd_1} and we follow the mappings described there to factorize it .

\item
${\rm Xbox}_{b3}$: ${\rm Xbox}_b$ where $x_3>x_{1,2,4}$ 
has the singularity structure of eq.~\ref{example4} and we use the mapping of eq.~\ref{mapping4}.

\item
${\rm Xbox}_{b4}$: ${\rm Xbox}_b$ where $x_4>x_{1,2,3}$ 
has the singularity structure of eq.~\ref{example2} and we use the mapping of eq.~\ref{mapping2}.
\end{enumerate}

We end up with 18 integrals to be evaluated numerically. This should be contrasted with the 119 sectors that are necessary if one factorizes the non-planar double box with sector decomposition. 


\section{Conclusions}
\label{sec:conclusions}

Higher order perturbative calculations are  very important for precision 
phenomenology at  modern accelerator experiments.  We believe  that NNLO computations will be particularly relevant  for 
signals  of  yet undiscovered  physics, such as a Higgs boson or  candidates of dark matter, in $2 \to 1$ and $2 \to 2$ processes. 
This motivates  the development of  powerful integration methods of matrix-elements  of up to  two virtual or real,  potentially unresolved, 
 partons.  Such integrations entail the disentanglement of overlapping singularities. 

In this paper, we  have introduced a method for the factorization of singularities  based on non-linear transformations. 
As proof  of principle, we presented the most singular integral topologies  which appear  in NNLO double-real 
radiation processes with massive particles in the final state. We  find that all overlapping singularities  can be  factorized  with our method, 
which yields  a small number of numerically  stable integrals. 
We have also applied our method to complicated crossed 
two-loop master integrals for massless QCD    scattering processes.  We  find that  we  can factorize  most of the 
overlapping  singularities with non-linear transformations. However, some remaining singularities  are  cumbersome  
to be treated  purely with our  method. In such situations, we  employ a hybrid  of our method and sector 
decomposition. This is more efficient  than  employing a pure sector  decomposition approach. 

The reduction of the number  of integrals which emerge in higher  order corrections  should facilitate NNLO computations. 
We  are looking forward to applying our  method for precision phenomenological studies  of  basic collider processes.

\section*{Acknowledgments}
We thank Andrea Banfi for many useful discussions  and private  communications of unpublished results. 
We thank G\"{u}nther Dissertori and Zoltan Kunszt  for discussions and motivation. 
This research is supported by the Swiss National Science Foundation under contract   SNF 200020-126632.




\bibliographystyle{JHEP}

\end{document}